\documentclass[aps,pre,twocolumn,superscriptaddress, nofootinbib]{revtex4-1}
\usepackage{graphicx}
\usepackage{amsmath,amssymb}
\usepackage[utf8]{inputenc}   
\usepackage{color}
\usepackage{booktabs}
\usepackage{multirow}

\usepackage[T1]{fontenc}
\usepackage{graphicx}  
\usepackage[caption=false]{subfig} 
\usepackage{mathtools}
\usepackage{breqn} 
\usepackage[titletoc]{appendix}
\usepackage{color}

\usepackage[normalem]{ulem} 

\definecolor{ao}{rgb}{0.0, 0.5, 0.0}

\newcommand*{\npart}{N_{\scriptscriptstyle\rm par}}   
\newcommand*{\nparteff}{N_{\scriptscriptstyle\rm par}^{\scriptscriptstyle\rm eff}}  
\newcommand*{\ndep}{N_{\scriptscriptstyle\rm d}} 
\newcommand*{\nrollcall}{N_{\scriptscriptstyle\rm rc}} 
\newcommand*{\Aij}{A_{\scriptscriptstyle\rm {d_i,  d_j}}}   
  \newcommand*{\depi}{\rm {d_i}}
\newcommand*{\depj}{\rm {d_j}}
\newcommand*{\q}{\rm {q}}

\newcommand{\be}{\begin{eqnarray}}
\newcommand{\ee}{\end{eqnarray}}
\newcommand{\ceff}{C^{\scriptscriptstyle\rm Eff}}
\newcommand{\cpart}{C^{\scriptscriptstyle\rm Party}}

\begin{document}
\title{On the stability of the Brazilian presidential regime: a statistical analysis}
\author{Frederico Fetter}
\author{Daniel Gamermann}
\author{Carolina Brito}
\affiliation{Department of Physics - Universidade Federal do Rio Grande do Sul (UFRGS) - Porto Alegre - RS - Brasil.}
\email{carolina.brito@ufrgs.br}


\begin{abstract}
Brazil's presidential system is characterized by the existence of many political parties that are elected for the Chamber of Deputies and unite in legislative coalitions to form a majority. Since the re-democratization in 1985, Brazil has had 8 direct presidential elections, among which there were two impeachments of the elected president. 
In this work we characterize the stability of the presidential regime and the periods of rupture analysing the votes that took place in the Chamber of Deputies from 1991 to 2019. 
We start by measuring the cohesion of the parties and the congress in the votes, quantifying the agreement between the votes of congressmen and observe that there is a stronger polarization among congressmen during legislative periods where there was no impeachment, referred to here as stable legislative periods. Using clustering algorithms, we are able to associate these polarized groups observed  during the stable periods with the opposition to the government and government base.
To characterize the impeachment of Dilma Roussef that happened in 2016 we analyze how the agreement between congressmen and the government evolved over time and identified, using cluster algorithms, that all the parties belonging to the majority coalition of the president, except her own party and another one, migrated to the opposition just before the impeachment.
Our analyses allow us to identify some differences between stable presidential periods and Legislative terms with an impeachment.
\end{abstract}


\maketitle


\section{Introduction}

Modern day technologies allow for the easy gather, storage and distribution of data relevant in many fields, such as political sciences, which have been traditionally tackled mainly by qualitative approaches. Quantitative analysis of these data should give insights in the workings of the many democratic institutions that run most of the countries in the world. The idea behind western-style representative democracy \cite{democracy} is that, through elections, the people choose those that better represent their interests for these many democratic institutions and that the plurality of these institutions guarantees the checks and balances needed to prevent abuses and self interests from prevailing. Moreover, representatives with common goals and ideals gather together to form political parties which, in principle, ease the identification of the electorate's interests within these groups of representatives.

One of such institutions is the chamber of representatives or national congress, where laws are analysed and voted by the congressmen (elected representatives of the people). Several works analyse data from elections \cite{iran, stylized}, its financing and how inappropriate money gathering might threaten the idea that the elected candidates do represent the people's interests \cite{tam2007breaking, spoils, gamermann}. Also the dynamics in the legislative chambers is analysed in different contexts. Such works might take into account the contents of the issues discussed and voted by the congressmen \cite{idprop, fader, issue}.  Ideal point models are based on our intuition that congressmen have positions in an abstract ideological space and cast their votes in roll calls based on where the voted bills lie in this space \cite{poole, clinton}. Such approaches, though, require an assessment not only of the voting data, but also of the contents of bills and speeches, which is invariably biased by the readers interpretations and therefore not indisputably objective. Other approaches study only the similarities between the votes cast by the politicians \cite{ussenate, MarencoPlosOne2020}.

{In the present work, we follow this former approach and analyse the voting patterns in the Brazilian Chamber of Deputies (the lower chamber of Brazil's bicameral legislative power, composed by this chamber and the senate). Our approach is blind to the contents of the bills voted in the congress. It analyses the cohesion of the votes results for each bill and the similarities in the votes sequences of each pair of congressmen using the $k$-means clustering algorithm. The objective of this clustering analysis is to identify the {\it de facto} groups of congressmen that vote cohesively in the roll calls and some of the dynamics behind these groups. We note that there are more than 30 political parties represented in the Brazilian congress \cite{tse} and the president's party alone does not have majority in the legislative. Therefore, in order to have a stable government, the president resorts to what has come to be called ``coalitional presidentialism''~\cite{abranches1988, rethinking, coalitional, abranchesBook2018}. This concept is defined as a strategy of directly elected minority presidents to build durable, cross-party support in a multiparty presidential regime.  The nominal 30 different political parties end up split into fewer actually interacting groups \cite{Laakso} in the legislature. We also investigate how changes in this split might lead to an unstable government resulting in the impeachment of the president.}

{The data analysed, the methods and metrics used in the study are explained in the next section. In section III we present the results highlighting the differences between stable regimes and those where an impeachment happened. Finally we present a discussion, summary and our conclusions.}


\section{Data and Methods} \label{method}


\subsection{Legislative periods analyzed in this work}

Table~\ref{legislaturas_tab} presents  information about the eight legislative terms analyzed in this work. It shows the number of the legislative period, the  abbreviation of the name of the presidents and how we will refer to them along the paper, the  start and end date of the term of each president, the total number of congressmen $\ndep$ and the total number of roll calls $\nrollcall$ in the period. The names of the presidents are Fernando Collor de Mello (identified as Collor), who resigned as president in 1992 to prevent an impeachment. He was followed by Itamar Franco (Itamar), then Fernando Henrique Cardoso (FHC), who had two consecutives mandates (that we refer to as FHC I and FHC II). Luiz Inácio Lula da Silva (Lula) also had two mandates, followed by Dilma Rousseff (Dilma) who completed her first mandate (Dilma I) and was ellected for the second term (Dilma II) but was removed from office through an impeachment process in 2016 and replaced by Michael Temer (Temer). In 2019 Jair Messias Bolsonaro (Bolsonaro) became president of Brazil. A normal legislative term in Brazil lasts for 4 years but in the case of Bolsonaro we will show the data of the first year of mandate (until Dec/2019).  The table also indicates the total number of parties $\npart$ elected to the Chamber of Deputies \cite{BancadasNaPosse} and an effective number of parties  $\nparteff$ \cite{Laakso1979}, which takes into account the number of congressmen per party and gives a better idea of the fragmentation of the congress.  This effective number of parties is defined as: $\nparteff = 1/{\sum\limits_{i} p_i^2}$, where $p_i$ is the proportion of seats the $i^{th}$ party has.
The $\nparteff$ is equal to the actual number of parties if every party has the same size and is closer to one if most congressmen belong to only one party. These numbers are calculated  with the parties assignments in the moment of the elections results.

\begin{table}
\caption{A summary of the Legislative periods analyzed in this work. The columns show respectively: the Legislature number, the abbreviation of the name of the president, the starting and ending date of the term, number of congressmen $\ndep$ and number of roll calls $\nrollcall$ 
voted during the respective legislative period.}
\label{legislaturas_tab}
\centering
\begin{tabular}{c | l | l | l|l|l|l|l}
      \multirow{2}{*}{Legis.} & {\scriptsize Presidential}   & \multirow{2}{*}{Start date}    & \multirow{2}{*}{End date}                 & \multirow{2}{*}{$\ndep$} & \multirow{2}{*}{$\nrollcall$} & \multirow{2}{*}{$\npart$} & \multirow{2}{*}{$\nparteff$}   \\   
      & \small{\scriptsize Term}   &     &                 & &  &  &    \\ 
      \hline \hline 
       \multirow{2}{*}{49}               & Collor  & 01/02/1991   &  28/09/1992  & 507 & 73   & \multirow{2}{*}{19} & \multirow{2}{*}{9.1} \\\ 
                                 & Itamar  & 29/09/1992   &  31/01/1995  & 506 & 85   & &  \\\hline 
                             50  & FHC I   & 01/02/1995   &  31/01/1999  & 590 & 468   & 18 &  8.1 \\\hline 
                             51  & FHC II  & 01/02/1999   &  31/01/2003  & 583 & 419   & 17 &  7.1  \\\hline 
                             51  & Lula I  & 01/02/2003   &  31/01/2007  & 566 & 450   & 18 &  8.4 \\\hline 
                             53  & Lula II & 01/02/2007   &  31/01/2011  & 559 & 611  &  20 &   9.3  \\\hline 
                             54  & Dilma I & 01/02/2011   &  31/01/2015  & 583 & 430   & 22 & 10.4 \\\hline 
       \multirow{2}{*}{55}                  & Dilma II & 01/02/2015   &  11/05/2016 & 553 & 330  & \multirow{2}{*}{28} &  \multirow{2}{*}{13.4}  \\
                                 & Temer    & 12/05/2016   &  31/01/2019 & 570 & 525  & &  \\ \hline
                             56  & Bolsonaro & 01/02/2019   &  31/12/2019    & 548 & 329  & 30 & 16.4  \\\hline 
\end{tabular}
\end{table}

We close this section with a remark about the ending dates shown in Table~\ref{legislaturas_tab}. In the legislative terms where there was an impeachment (Collor-Itamar  and Dilma-Temer), the ending date corresponds to the moment at which the president was removed form her/his position. According to Brazilian law, the president is removed from office once the impeachment process is started and the official judgment of the impeachment happens three months latter by the Senate, when she/he might be brought back if declared not guilty or be officially removed otherwise.


\subsection{Data format}

Brazil has excellent transparency laws which have been very well implemented and nowadays it is fairly easy to access huge amounts of data concerning the public administration and the legislative \cite{dadosabertos}. In this work, we use data available from the Brazilian Chamber of Deputies, concerning roll calls in the national congress \cite{camaraori, camara}. From the application programming interface (API) developed by the I.T. personal working in the congress, one can obtain a list of roll calls voted in a given year and the votes cast by the congressman in each open roll call.

In  Table~\ref{dados2} we show schematically the type of data we obtain from this database. Each roll call is represented in Table~\ref{dados2} by $r^q$, where $\q=1,..,\nrollcall$ where $\nrollcall$ is the total number of roll calls voted during a given legislative period. The congressmen are represented by $\depi$, where $i=1,..,\ndep$ where $\ndep$ is the total number of congressmen.
For each congressman $\depi$ one has a sequence of votes which can be represented by  a vector of options ${\bf o}_i = (o_i^1, o_i^2, \dots, o_i^{\nrollcall})$ where each $o_i^q=v_j$ can assume 5 different values:  $v_1=$Yes, $v_2=$No, $v_3=$Abstention, $v_4=$Obstruction and $v_5=$Art.17\footnote{This last option is reserved for a small fraction of congressmen that compose the presiding table at the plenary.} (a given congressman may also be absent from a given roll call, an issue which will be discussed latter).

For each congressman, there is the information about his party affiliation and the federal unity she/he represents.
Some roll calls also contain information about how the government, coalitions and the parties oriented their congressmen to vote.

\begin{table}
\caption{Schema of structure of the data. Each roll call is represented by $r^k$ and congressmen are represented by $d_i$. For each deputy, we know a list of he/her votes in each roll call: ${\bf o}_i = (o_i^1, o_i^2, \dots, o_i^{\nrollcall})$ and $o_i$ may have 6 different options, as explained in the text. For each deputy there are also other information as for example the party  $p_{i}$ to which he/she belongs.}
\label{dados2}
\centering
\begin{tabular}{c|cccc|c}
       deputy           &  $r^1$                &    $r^2$        & \hspace{3mm}... \hspace{3mm}   &        $r^{N_p}$          &   party                 \\ \hline \hline 
       $ {\rm d}_1$        	   &  $o_{1}^{1}$           &   $o_{1}^{2}$   & ...                                                  &       $o_{1}^{N_p}$      &    $p_1$                 \\ \hline
       ${\rm d}_2$                  &  $o_{2}^{1}$           &   $o_{2}^{2}$   & ...                                                  &       $o_{2}^{N_p}$      &   $p_2$                  \\ \hline
      $\vdots$               &  $\vdots$           &   $\vdots$     & $\ddots$                                                  &       $\vdots$       &  $\dots$                 \\ \hline
       ${\rm d}_{\ndep}$                  & $o_{\ndep}^{1}$          &   $o_{\ndep}^{2}$   & ...                                                  &       $o_{\ndep}^{N_p}$      &  $p_{\ndep}$                  \\ \hline
\end{tabular}
\end{table}

It is important to point out some characteristics of these data to understand the limitations of our study.
The roll calls whose votes are registered in the database \cite{camaraori, camaraimprensa} are called ``nominal'' or open. These are the ones we evaluate and they represent less than 20\% of the total vote sessions in the congress;  more than 80\% of the votes in the plenary are secret and only the results are made public\footnote{From all roll calls voted per year that one can download from \cite{camara}, only in around 20\% of the cases one is able to obtain the list of votes by congressman. The real number of secret and open roll calls might be different since the database itself claims not to be completely up to date.}. Moreover most of the bills are never put to vote and this is related to the fact that mostly, it is the President of the Chamber of Deputies who decides which bills are voted and in which order. This choice is clearly not random, but is  subject to political calculation. This introduces a bias in our analysis that is a common feature of many roll call analysis studies \cite{Figueiredo2000}.

We end this section with three considerations about our choice of analyses. In this work we consider all roll calls of the database as equally important, without giving weight to different types of projects or the subject tackled in the bills and roll calls. Considering the number of congressmen, there are 513 congressmen elected for each legislative term. Some of them are nominated as ministers or any other functions and are replaced by an alternate congressman. Then, in practice, there are much more than 513 congressmen who vote along the four years of mandate. Some congressmen participate in very few roll calls and we exclude them from our analysis with a criterion that is explained in  the Appendix \ref{append_a}. This question of congressmen migration  and participation in the roll calls is studied in detail in~\cite{Vieira2019}. Also, congressmen can change parties during a legislative period. In this work we assume that a deputy belongs to the party for which he voted differently from ``Absent'' for the first time.


\subsection{Measurements}

In this subsection we define the quantities used in this work to analyse the legislative data.


\subsubsection{Cohesion in the roll calls} \label{cohesionparams}

{As mentioned above, there are 5 different ways a congressman may vote besides being absent in a roll call. These different alternatives do point to different strategies the government, opposition and the different parties may be adopting given the bill at hand. There are of course the yes and no alternatives indicating support to approve or dismiss a given bill, but other alternatives like obstruction or abstention may indicate a push to either postpone or alter the bill.}

{In order to evaluate the cohesion of a given party in a roll call, and therefore to assess whether the party is cohesively following a given strategy, we adopt a concept from information theory: the Shannon entropy \cite{shannon}, which measures the uncertainty in a distribution. Given that a fraction $p_i$ of the congressman from a given party voted option $v_i$ in a given roll call, one can evaluate}

\be
S &=& -\sum_{i=1}^{n_v} p_i\ln p_i \\
C &=& 1-\frac{S}{\ln n_v},
\ee
{where $S$ is the Shannon entropy for the distribution of $p_i$'s, which assumes values between 0, when all congressmen in the party voted the same option (zero uncertainty), and $\ln n_v$ when the congressmen voted evenly among the $n_v$ different options voted (maximum uncertainty). Note that the value of $C$, which we will be calling cohesion, will be between 1 (cohesive strategy adopted) and 0 (party most divided).}

{Given a roll call, one can evaluate two global cohesions associated to it, which are going to be called effective cohesion $\ceff$ and party cohesion $\cpart$. The effective cohesion is evaluated using the $p_i$ distribution in a given roll call without regard to the parties, just evaluating the total number of congressmen that voted each option. For evaluating the party cohesion, given a roll call, first for each party the particular party cohesion is evaluated, and the total roll call party cohesion is then the weighted average of the parties cohesions:}

\be
\cpart &=& \frac{1}{N}\sum_j C_i n_{i}, 
\ee
{where $N$ is the total number of congressmen not absent from the roll call vote, the sum runs over all parties that participated in the roll call, $C_i$ is the cohesion of party $i$ in the roll call and $n_{i}$ is the number of congressmen from party $i$ that participated in roll call.}


\subsubsection{Agreement between congressmen}
\label{defAij}

To quantify  how similar are the sequences of votes between two congressmen $\depi$ and $\depj$, we define the agreement as:

\begin{equation}
   \Aij =  \frac{1}{N}\sum_{k}^{N} \delta^{\scriptscriptstyle\rm {q}}_{\scriptscriptstyle\rm {d_{i}, d_{j}}} 
\end{equation}
where $\q$ is the index of the roll call, $\delta^{\scriptscriptstyle\rm {q}}_{\scriptscriptstyle\rm {d_{i}, d_{j}}}$ is a Kronecker delta such that,  $\delta^{\scriptscriptstyle\rm {q}}_{\scriptscriptstyle\rm {d_{i}, d_{j}}}=1$ if congressmen $\depi$ and $\depj$ vote with the same option in the roll call $\q$ and $\delta^{\scriptscriptstyle\rm {q}}_{\scriptscriptstyle\rm {d_{i}, d_{j}}}=0$ otherwise, $N$ is the total number of roll calls that both congressmen $\depi$ and $\depj$ voted differently than Absent. With this definition, $\Aij=0$ if both congressmen voted completely different or if they are never present simultaneously in any roll call and  $\Aij=1$ if they have exactly the same sequence of votes in all roll calls they both participated in a given legislative period.


\subsubsection{$K$-means to identify groups in the Chamber of Deputies}
\label{kmeans}

In order to identify groups of congressmen who voted in a similar way, we use the $k$-means cluster algorithm \cite{elements}. Below we succinctly explain how this method works:

The point is to cluster $m$ observations lying in an $N$-dimensional space into $k$ clusters. The algorithm proceeds as follows:

\begin{enumerate}
\item randomly select $k$ observations. These will be called the centroids of the clusters

\item assign each observation to the cluster whose centroid is closest to it. 

\item update the centroids as the mean point of the observations belonging to the cluster
\end{enumerate}

Repeat the steps 2 and 3 until the the change in the centroids positions is bellow a given precision. This process will cluster the points which are closer together by minimizing the intra-cluster distance.

\begin{figure}[h!]
\centering
\includegraphics[width=0.46\columnwidth]{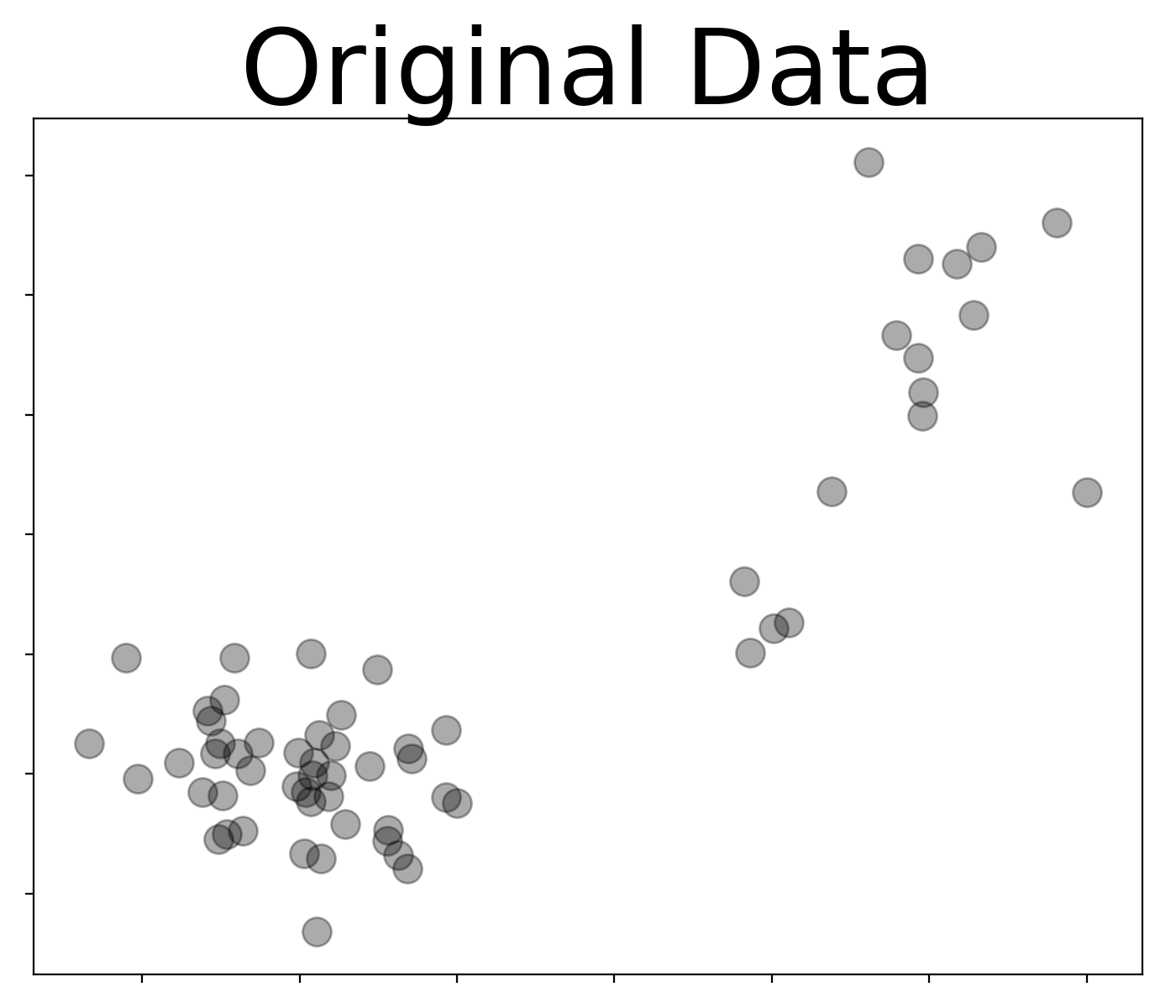} \\
\includegraphics[width=0.46\columnwidth]{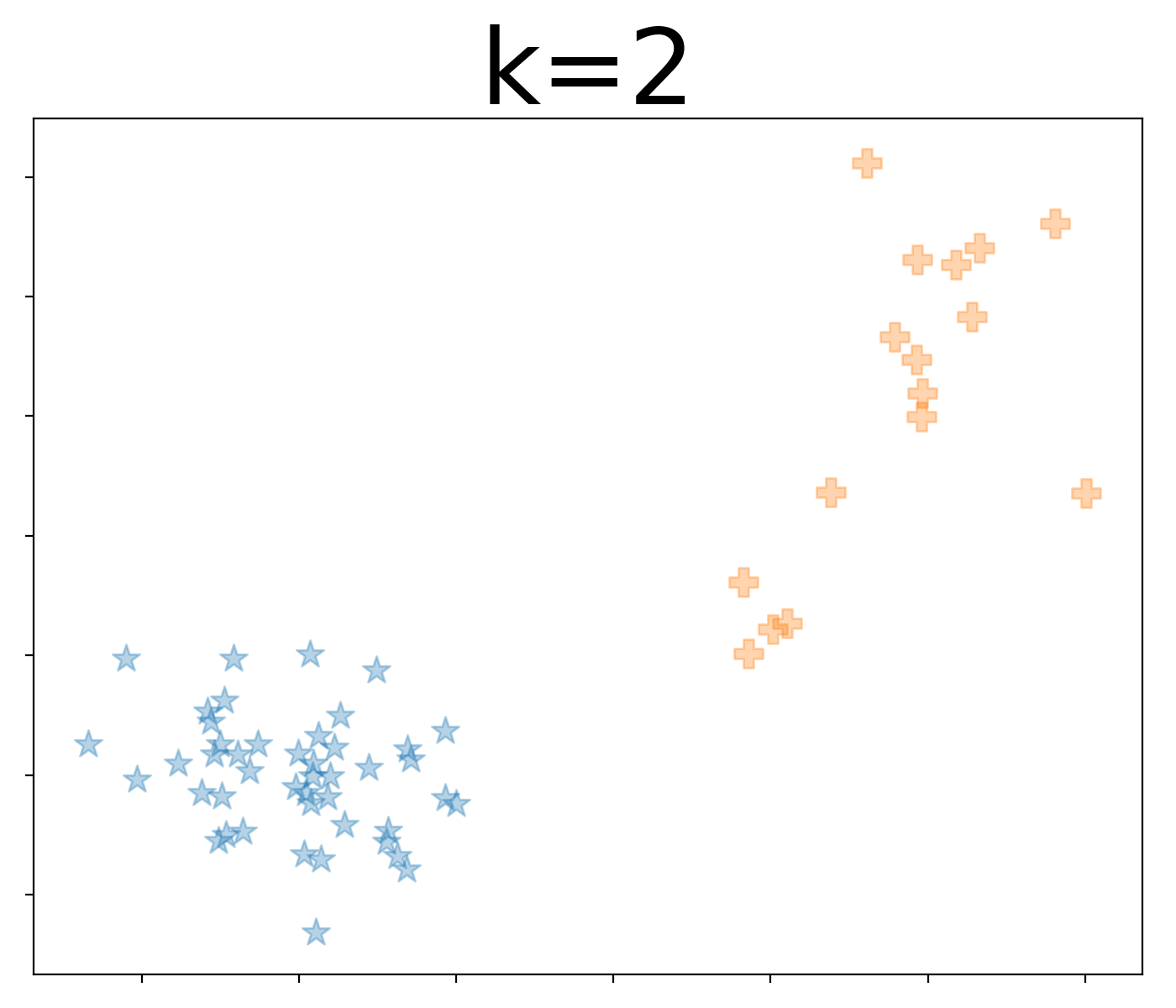}
\includegraphics[width=0.46\columnwidth]{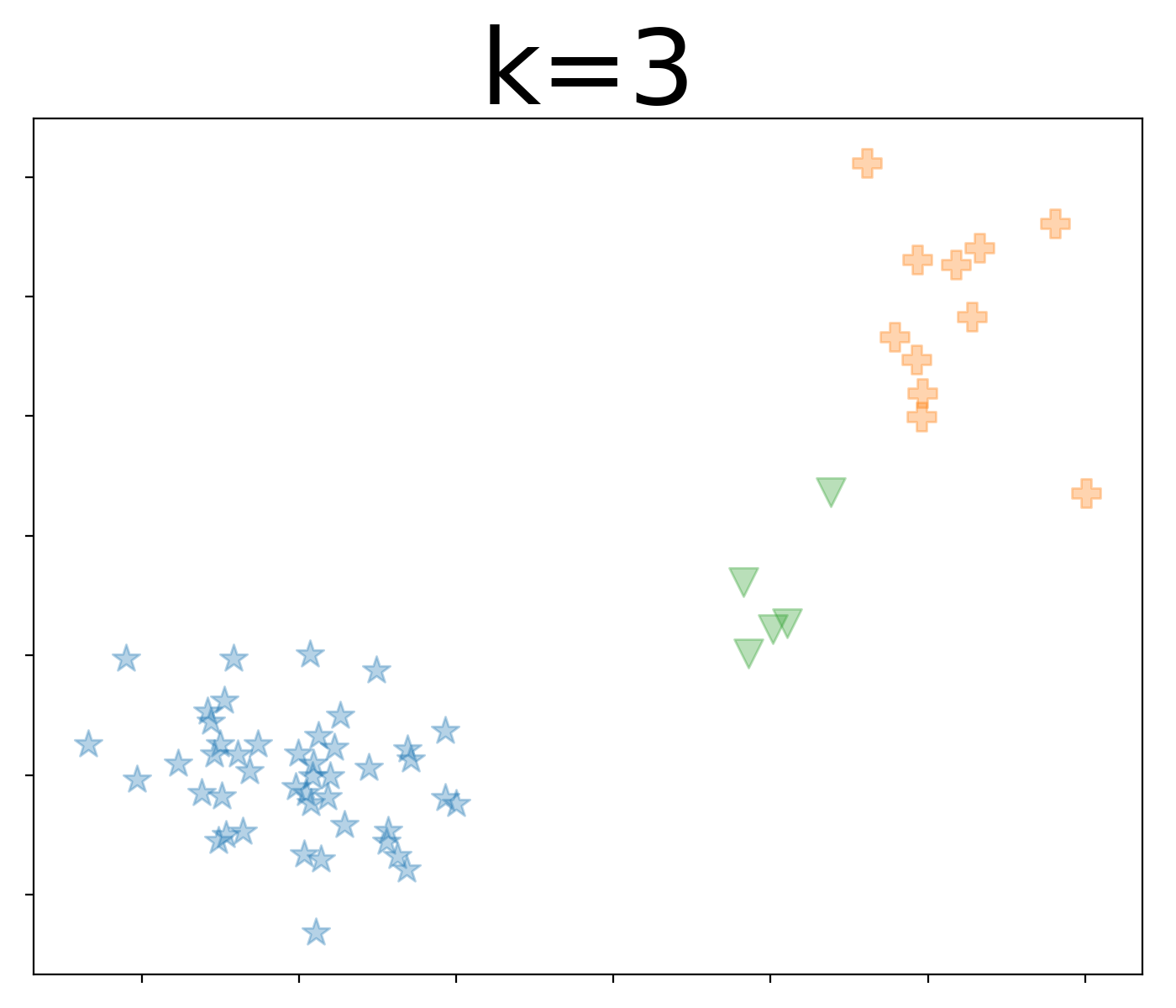}
\caption{Example of the $k$-means algorithm applied to 2-dimensional data. On the top it is shown the original data. In the lower left the points are clustered in $k=2$ groups, and identified by color and marker. In the lower right panel $k$-means was applied for $k=3$.}
\label{fig:kmeans}
\end{figure}

In our case, each observations will be the vector
$$
A_{i} = (A_{\mathrm{d}_i \mathrm{d}_1}, A_{\mathrm{d}_i \mathrm{d}_2}, \dots, A_{\mathrm{d}_i \mathrm{d}_{\ndep}})
$$
which characterizes the agreement of the $\depi^{th}$ congressman to all others. The absolute centroid position of each group is not relevant here, but each congressman has now a label indicating to which group he/she belongs i.e. to which other congressmen he is more afine. In Fig.~\ref{fig:kmeans} there is a pictorial representation of the results of the algorithm for a set of points.

In what follows, we refer to the groups identified by the this algorithm with a given $k$ as $G_n^{k}$, where $n=1, 2..., k$.

A caveat of this method is that the number of clusters is an input, not an output of the method. So, for the same data, one may run the algorithm with different values of $k$ obtaining different results. A criterion to assess the best number of cluster is the elbow method which seeks to identify the value of $k$ for which adding a new cluster does not significantly reduces the variance (also called in this case inertia) for the distribution of the data points around their assigned centroids \cite{elbow}.


\subsubsection{Alignment of the clusters with the president's party}
\label{Snk}

Once we define a group of congressmen $G^k_i$ using the $k$-means algorithm, we can quantify the degree of alignment or support of these congressmen with the president's party $P_p$ by defining the  following quantity:

\begin{equation}
  {\cal S}_n^k =  \frac{1}{|G_n^{k}||P_p|} \sum_{\depj \in G_{k}} \sum_{\depi \in P_p} \Aij,
\label{kappa}
\end{equation}
where $|G_n^{k}|$ and $|P_p|$ are the total number of congressmen in the group $G_n^k$  and in the president's party $P_p$, respectively. Note that it is just an average of the agreement over pairs of congressmen, one belonging to a cluster $G_n^k$ and the other to the presidents party $P_p$. This quantity is expected to be closer to 1 if the group $G_n^k$ has parties aligned politically with the  government and is expected to be smaller if $G_n^k$ has congressmen who belong to the opposition to the government.

The ${\cal S}_n^k$ can be defined in different intervals of time in a given legislative term  in order to study its change and behaviour along the time. The dynamics of this measure over time is a way to identify possible instabilities of a presidential term.


\section{Analysis of the legislative activity data}

In this section we present the results of the analyses using the data for the roll calls in the Chamber of Deputies.

\subsection{The roll calls in cohesion space}

As explained in section \ref{cohesionparams}, one can associate to each roll call two measures of cohesion, the effective cohesion $\ceff$ and the party cohesion $\cpart$. These cohesion parameters define, therefore, a two dimensional phase-space where one can locate each roll call.

\begin{figure}[h!]
\centering
\begin{tabular}{cc}
\includegraphics[width=0.35\columnwidth]{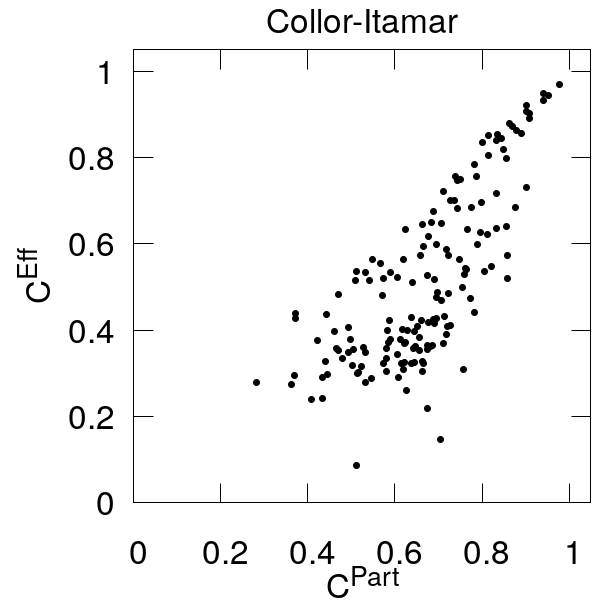} & \includegraphics[width=0.35\columnwidth]{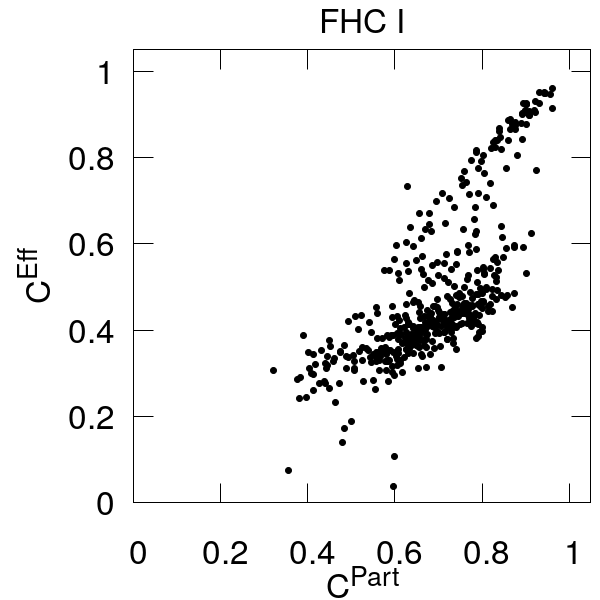} \\
\includegraphics[width=0.35\columnwidth]{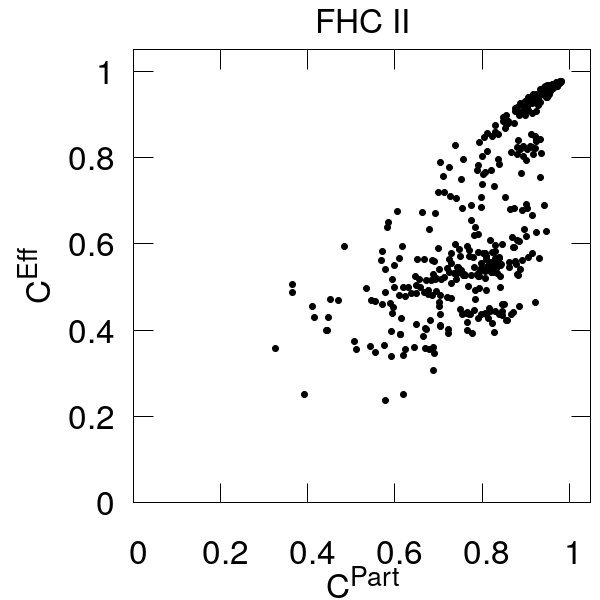} & \includegraphics[width=0.35\columnwidth]{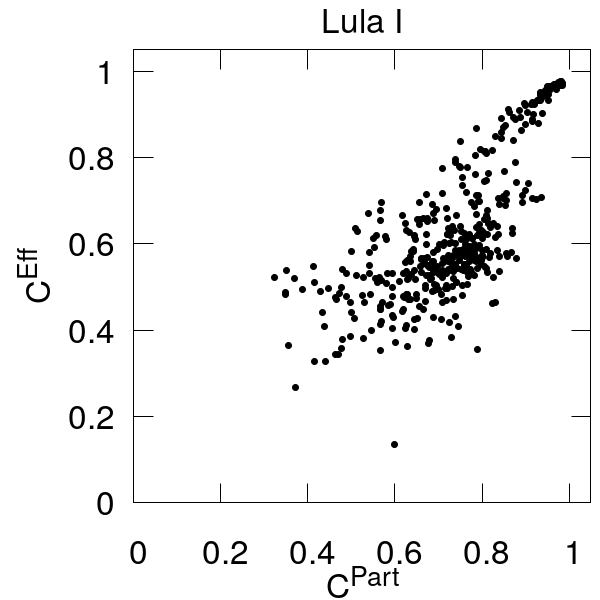} \\
\includegraphics[width=0.35\columnwidth]{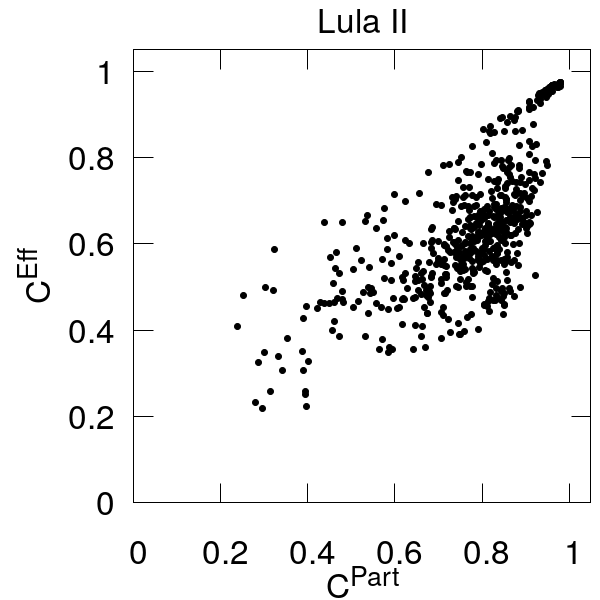} & \includegraphics[width=0.35\columnwidth]{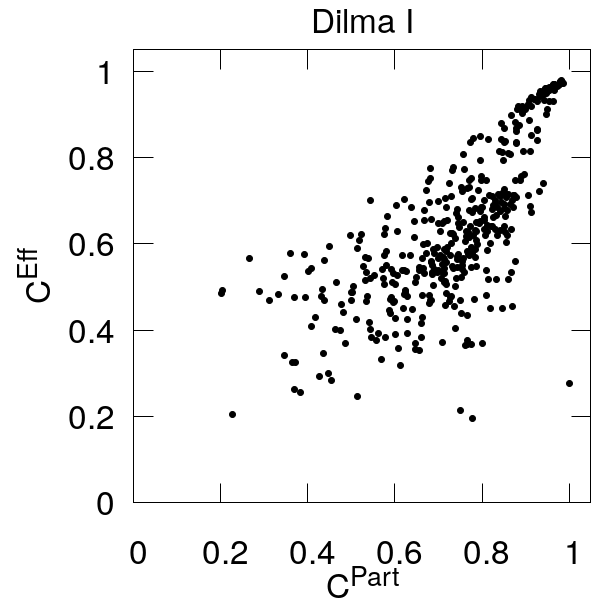} \\
\includegraphics[width=0.35\columnwidth]{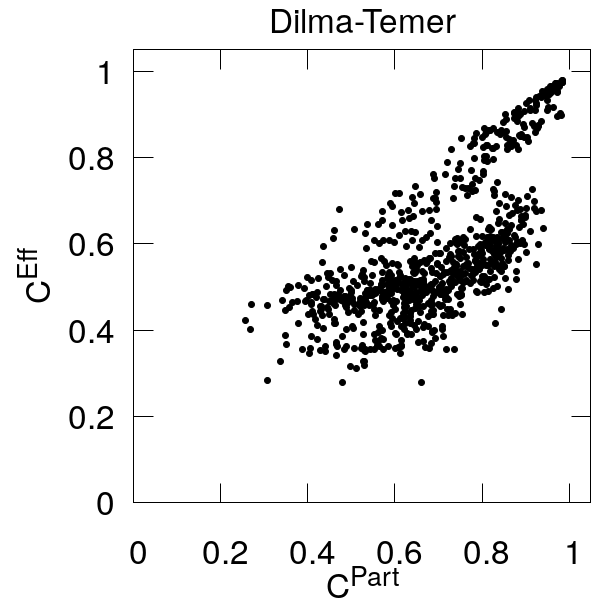} & \includegraphics[width=0.35\columnwidth]{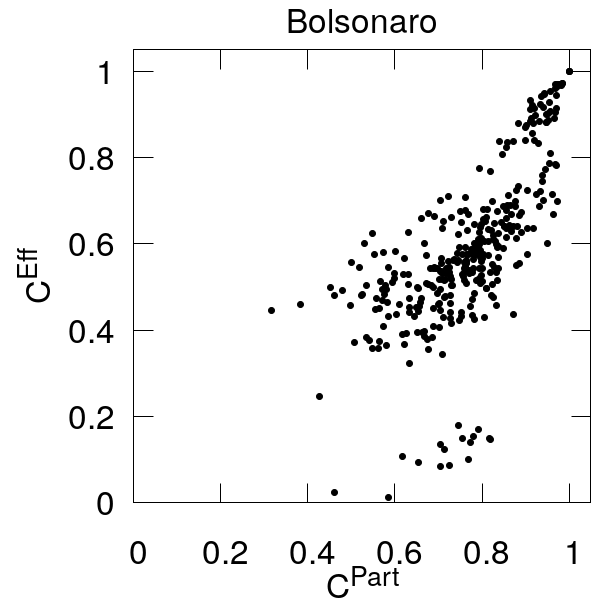} 
\end{tabular}
\includegraphics[width=0.80\columnwidth]{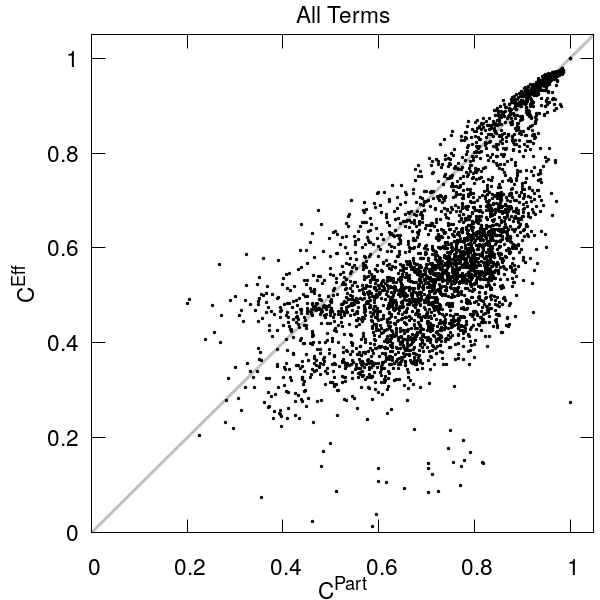}
\caption{Roll calls in cohesion phase space. The plots in the two columns show each legislative period separately and the last one below has all roll calls for all periods superposed. In this last plot, also the main diagonal line is traced in light grey color.}
\label{fig:votesinps}
\end{figure}

In Fig.\ref{fig:votesinps} it is shown the scatter plots of the roll calls for the different legislative terms in this cohesion phase space.  In these plots, each roll call is a point in the $\ceff\times\cpart$ space. Points on the diagonal line indicate that the cohesion inside a party is the same as the cohesion on the whole National Congress for a given roll call. Points below the diagonal indicate the roll calls that have higher cohesion inside of the parties than in the whole congress. From the figure it is possible to observe first that the same pattern repeats in all legislative periods: roughly, two groups of votes tend to form a more compact cluster, one with high cohesion and a more disperse one with sometimes high and sometimes low cohesion. Moreover, in either cluster of points, effective and party cohesion seem to be correlated, one tends to be high when the other is high as well. We should note here, that this correlation is not evident from the definitions of these parameters. One is measuring the parties intrinsic cohesion and the other the overall result of a roll call.  Were the congress evenly divided into two strongly opposed parties, a high party cohesion would imply a low effective cohesion for one party would systematically vote the opposite option than the other, meaning that the party cohesion would be high, but the effective cohesion low because the overall result of the vote would be divided. Therefore, it is interesting to observe here that very polarizing issues in the congress tend not only to unite intrinsically the parties, but also the parties among them and dividing issues, tend not necessarily to oppose parties against each other, but rather to divide the parties intrinsically.

Overall, these data suggest that there are two types of roll calls in terms of cohesion as we explain now. Some roll calls form a cluster on the top right in the scatter plots and lay on the diagonal line. These are roll calls for which there is a very high cohesion inside of the parties and in the congress itself. These roll calls correspond to subjects that are consensual and therefore are not useful to distinguish different ideologies among parties. On the other hand, more diffuse points represent roll calls for which there are less cohesion. Because these points appear for smaller values of C, they indicate that, both deputies inside a party or of the whole congress vote in a less  cohesive way. These can be seen as more controversial roll calls both among congressmen of a given party and among all the congressman of the National Congress.

\subsection{Distribution of the agreement between congressmen}
\label{sec:distAij}

\begin{figure}[h!]
\centering
\includegraphics[width=0.49\columnwidth]{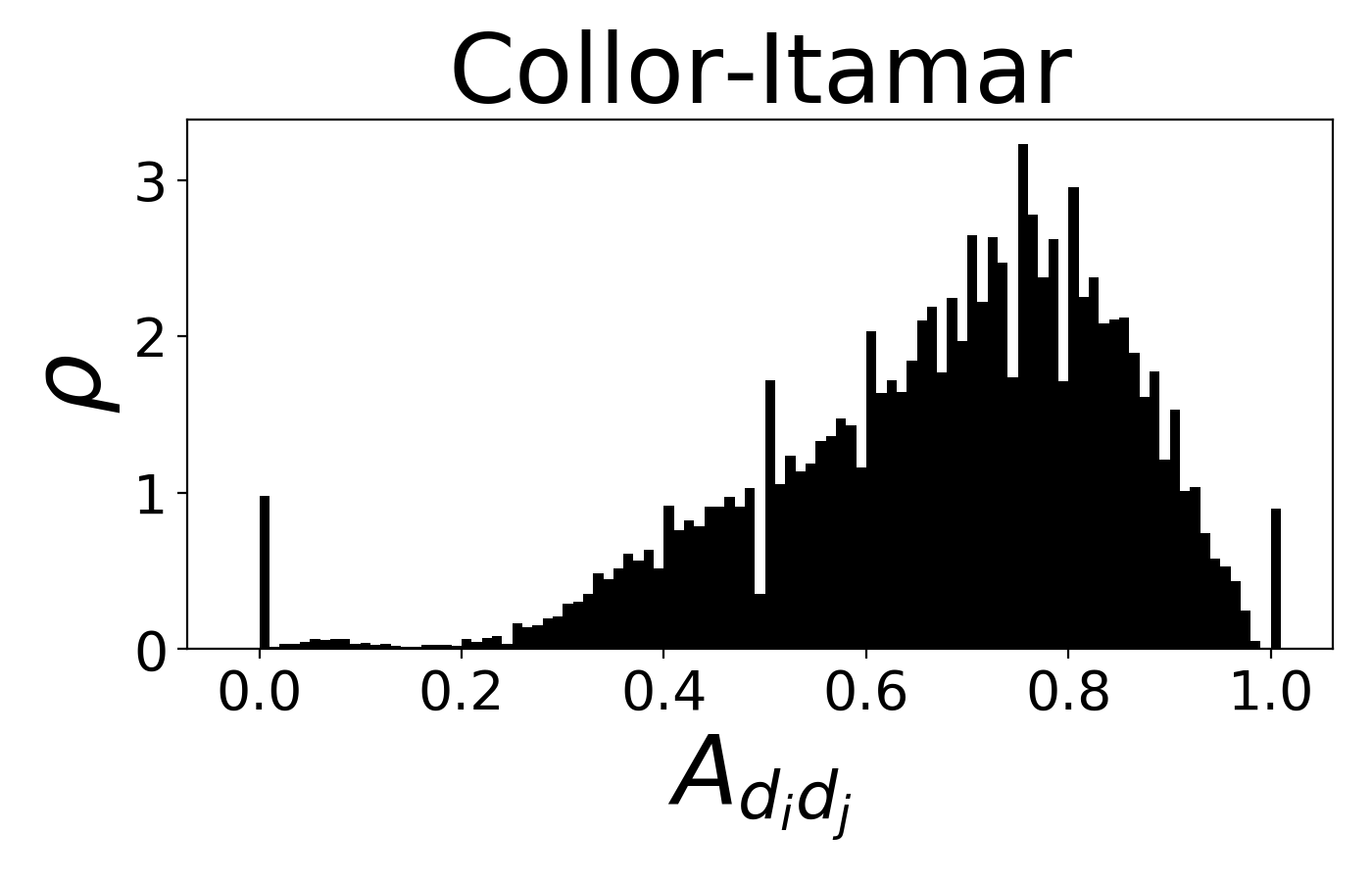}
\includegraphics[width=0.49\columnwidth]{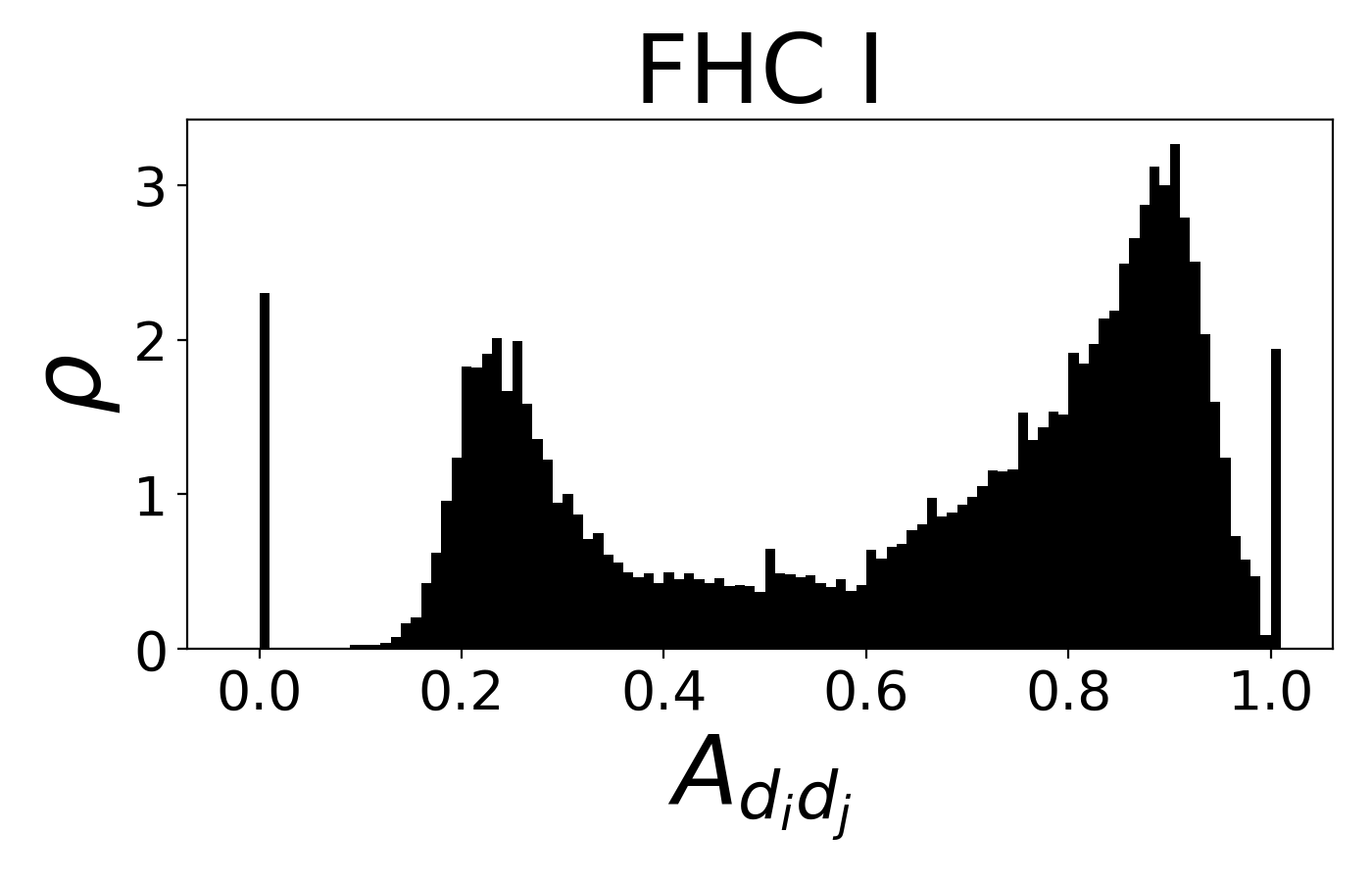}
\includegraphics[width=0.49\columnwidth]{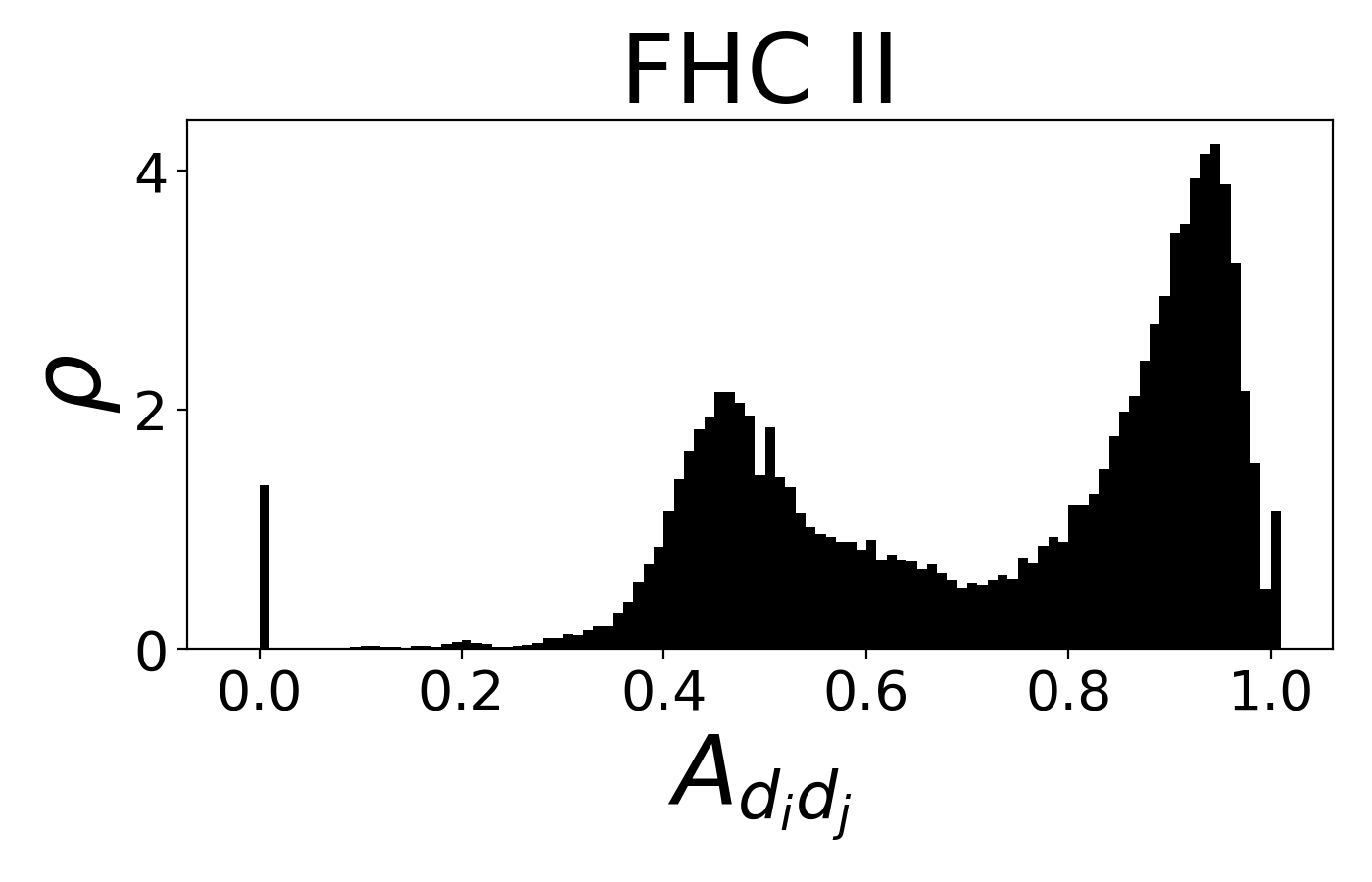}
\includegraphics[width=0.49\columnwidth]{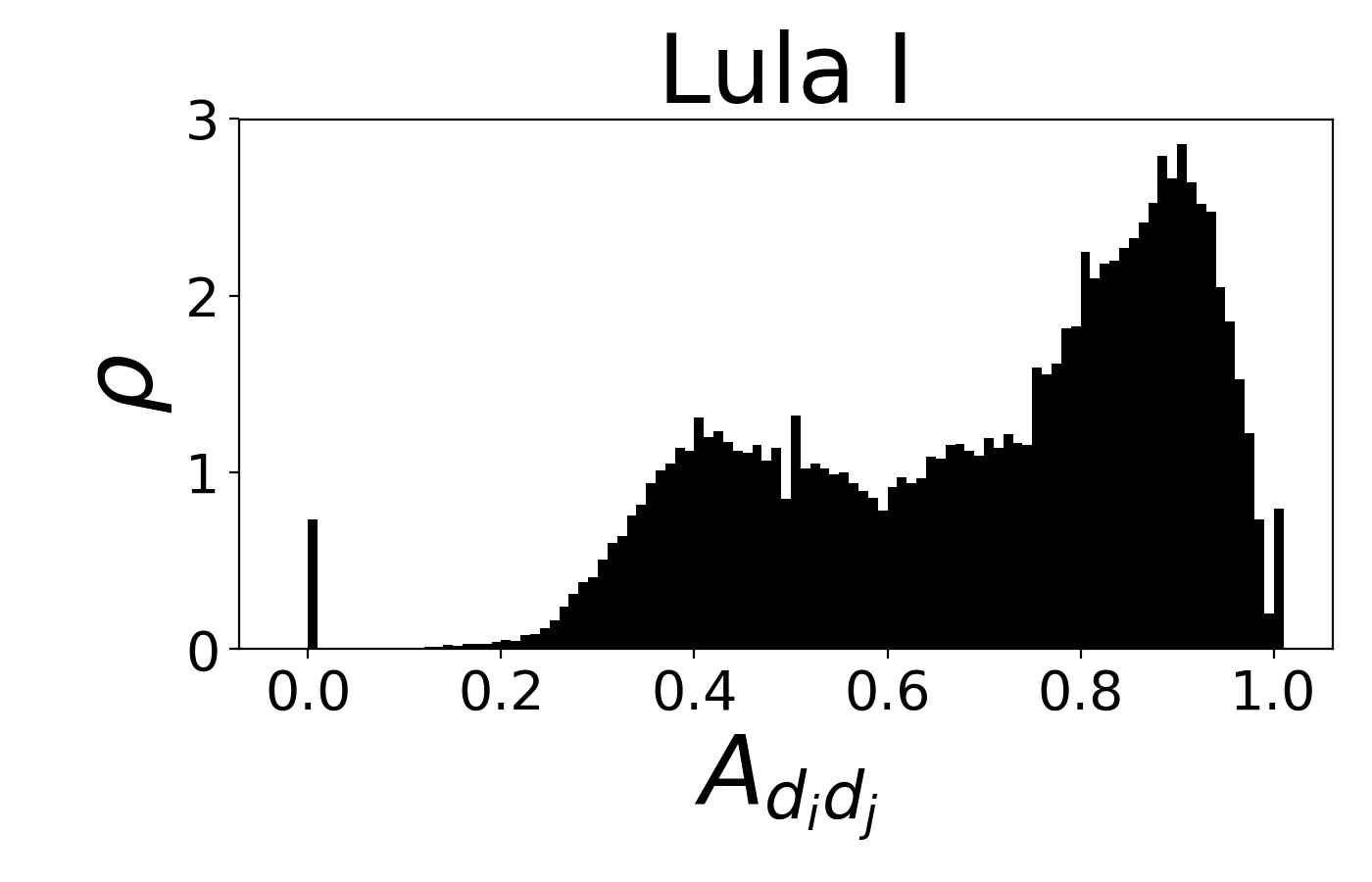}
\includegraphics[width=0.49\columnwidth]{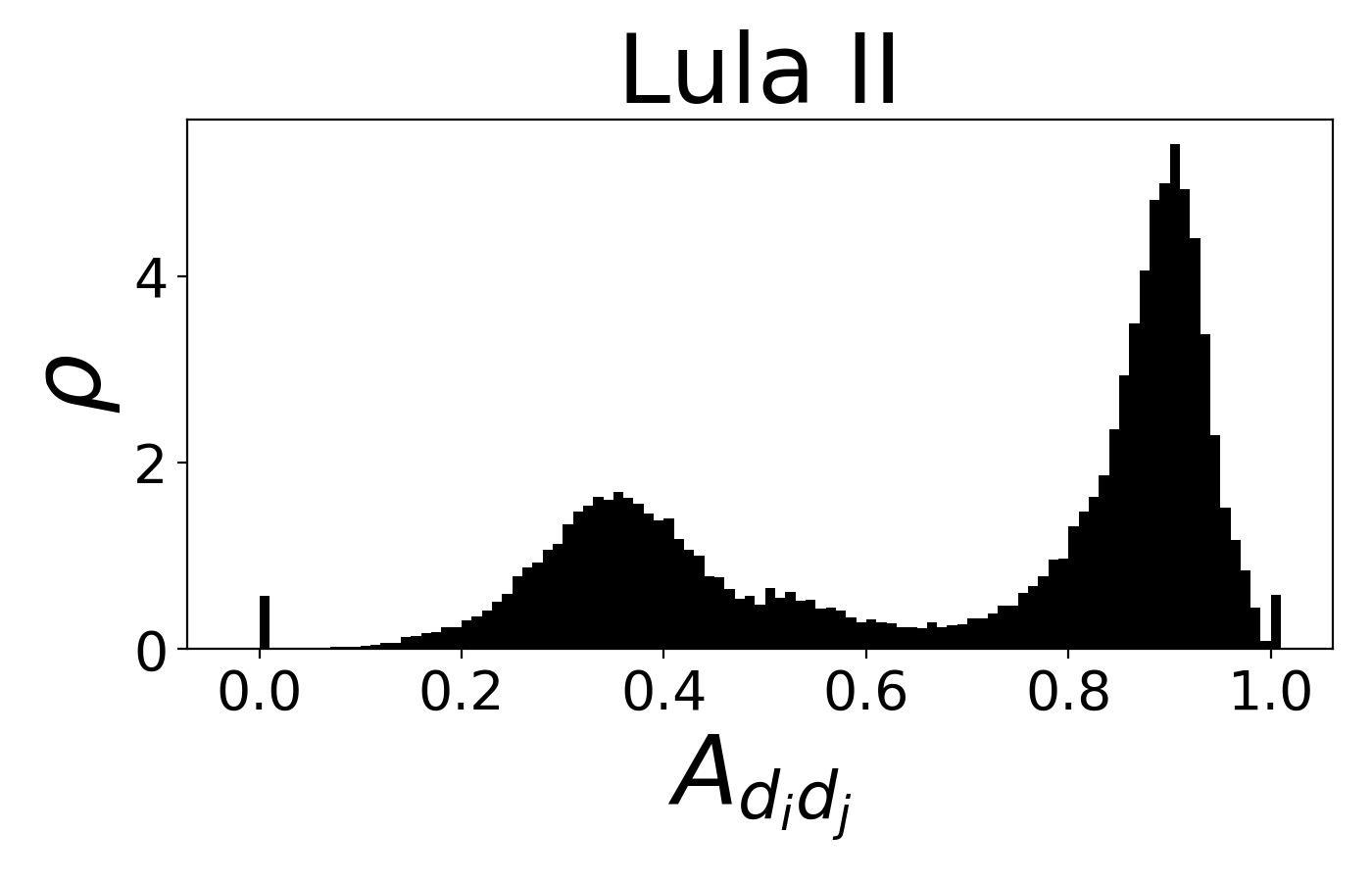}
\includegraphics[width=0.49\columnwidth]{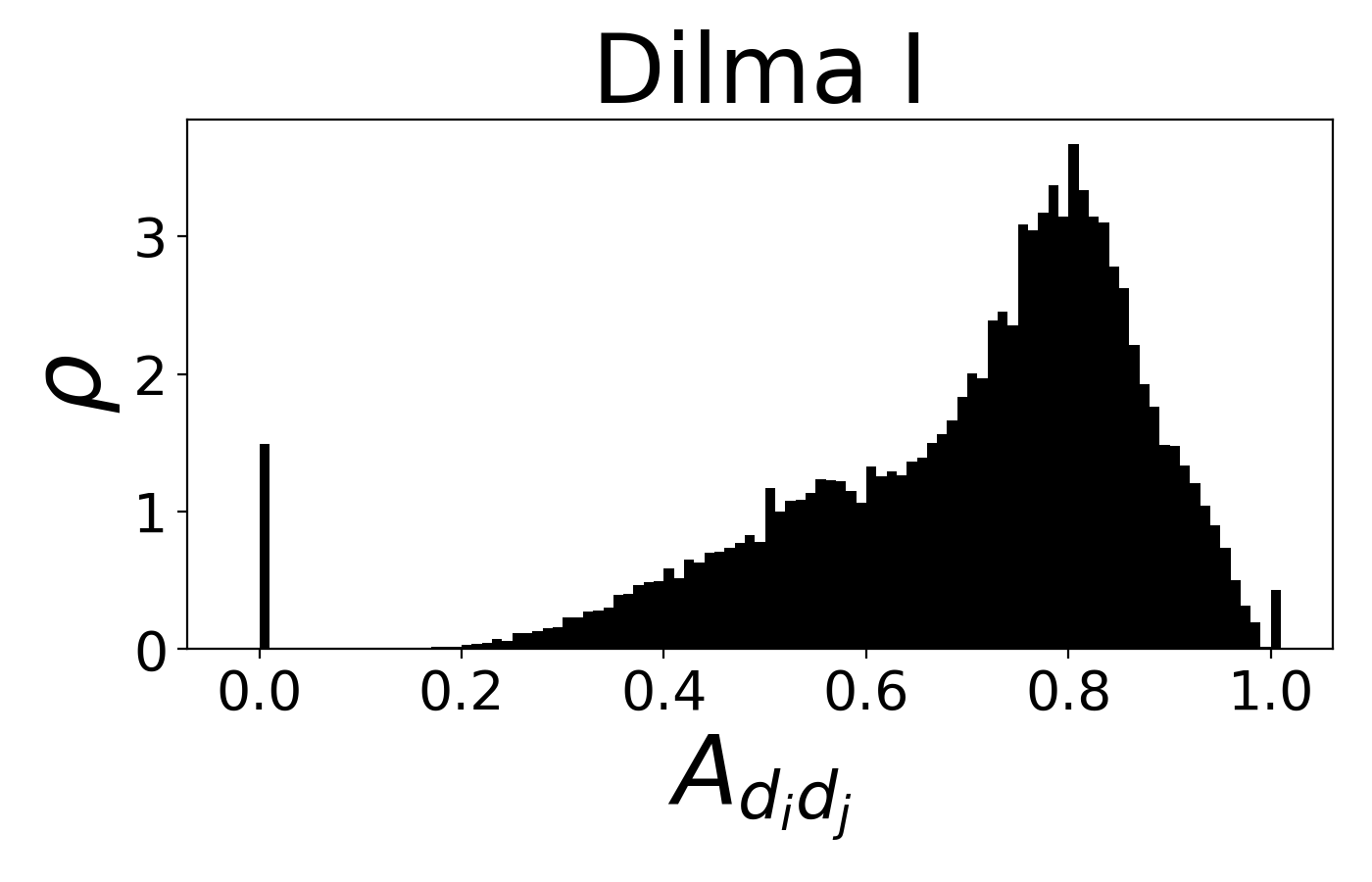}
\includegraphics[width=0.49\columnwidth]{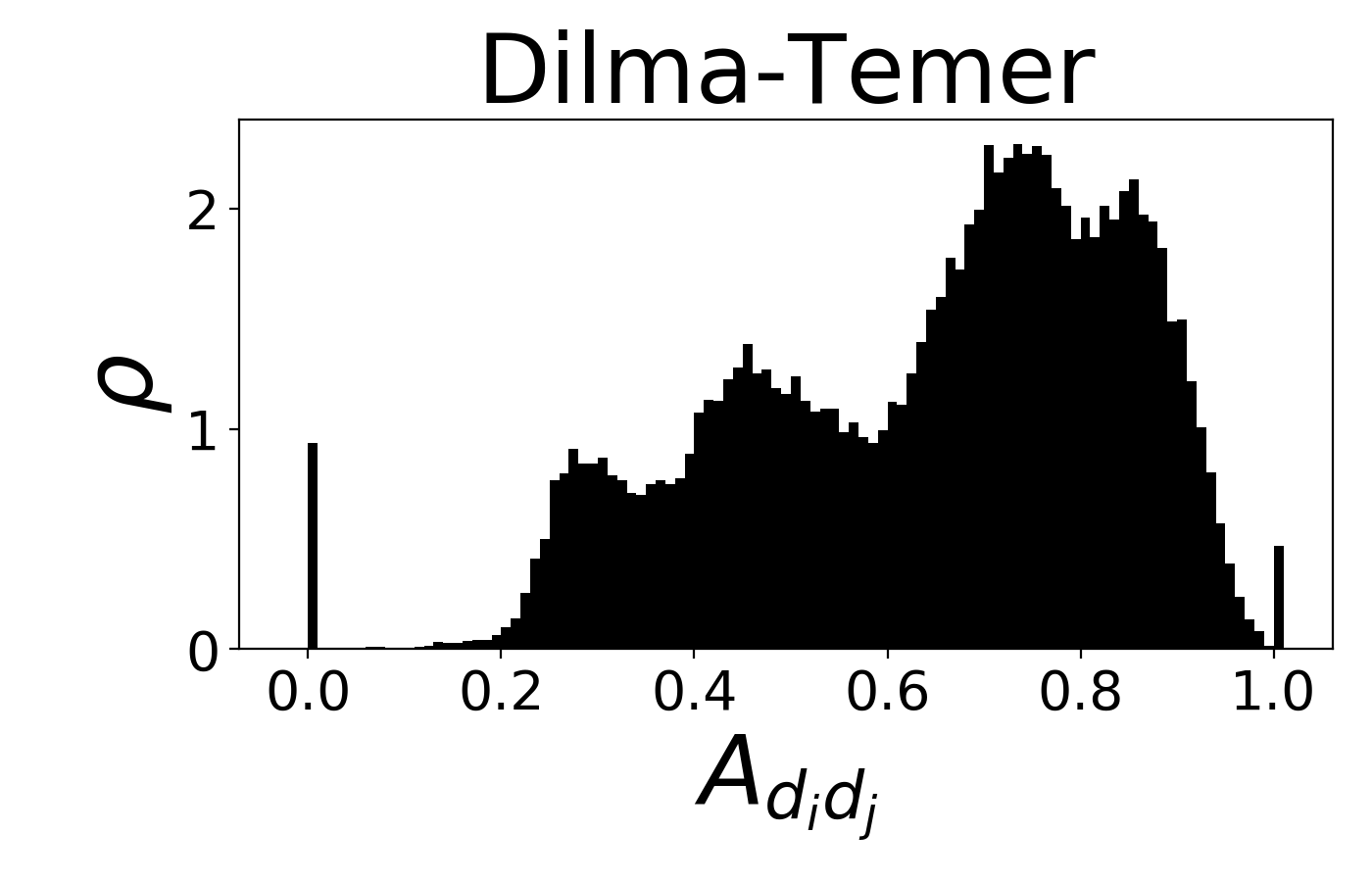}
\includegraphics[width=0.49\columnwidth]{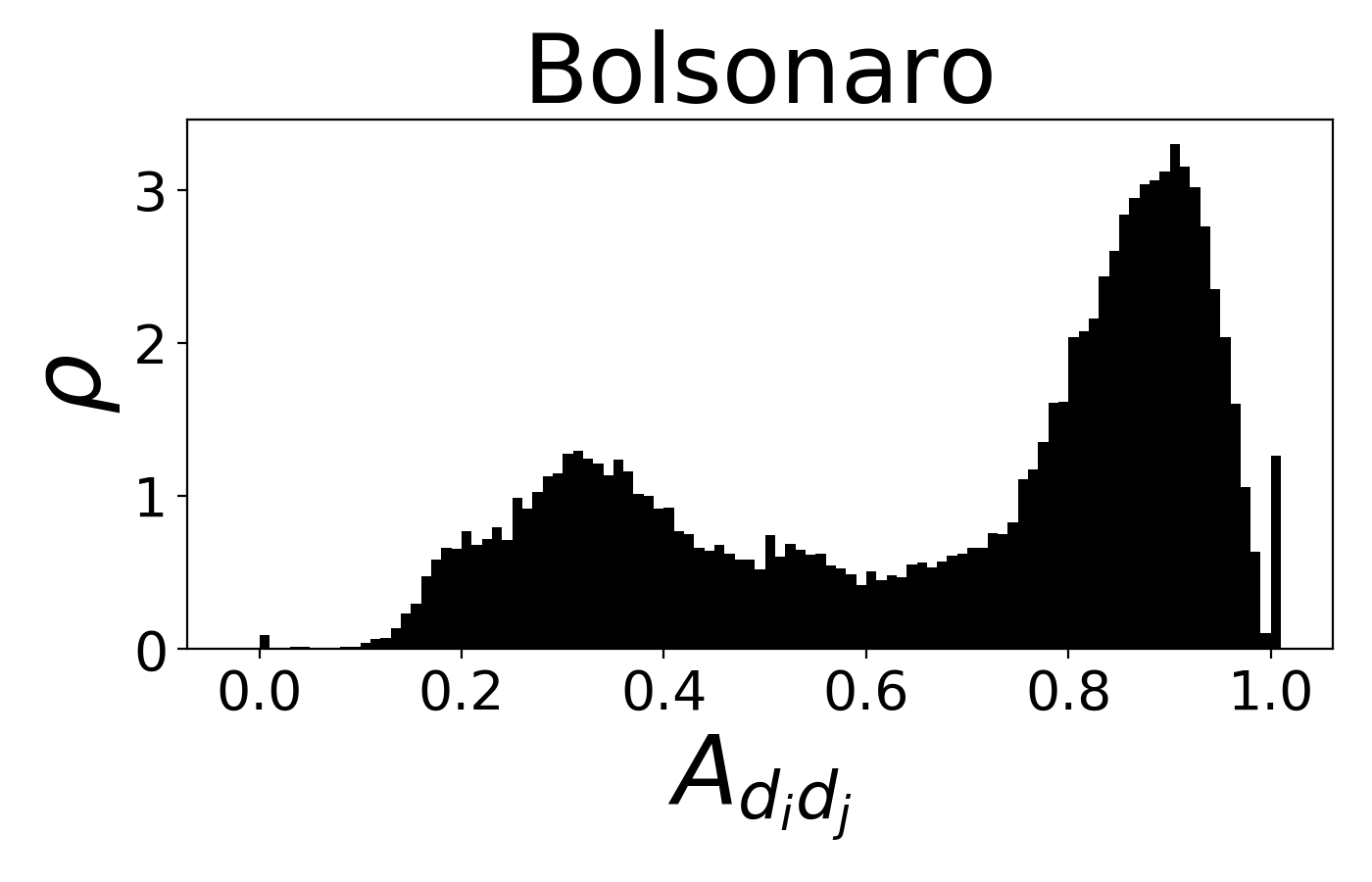}
\caption{Normalized distribution of the agreement matrix elements for each analyzed period.}
\label{fig:distribuicoesAij}
\end{figure}

Fig. \ref{fig:distribuicoesAij} shows the distribution of $\Aij$, as defined in section \ref{defAij}, for the eight legislative terms considered in this work. By inspecting the distributions we identify that most of the terms presents a bimodal distribution,  the exception being the periods governed by Collor-Itamar, Dilma I and Dilma-Temer.  A similar pattern was observed using a different type of measure of agreement \cite{MarencoPlosOne2020}, where the authors fit a bimodal function to the distributions and they find and associate the relative distance between the peaks to an indicative of instability. 

The periods  Collor-Itamar and Dilma-Temer are referred to as  {\it politically unstable} because in both there was  an impeachment of the president\footnote{Officially, Collor resigned before being impeached to avoid loosing his political rights (a move similar as done by Nixon).}.  When  Dilma-Temer term is separated in two parts, one before and the other after the impeachment, into a Dilma II period and a Temer period (as shown in Fig. \ref{fig:Aplicak2}),  it is observed that the agreements in the Dilma II period are unimodal while in Temer's are bimodal, very similar to FHC, Lula or Bolsonaro. The same process shows that both Collor and Itamar are separately unimodal.

This observed bimodality during stable legislative periods suggests that a polarization {into two opposing blocks seems} necessary to stabilize the Brazilian political system. This is a curious result because Brazil has one of most fragmented political system in the world, having many political parties, as shown in Table \ref{legislaturas_tab}. To better characterize these periods of stability and instabilities,  in the next sections we cluster congressmen in groups and analyse their behavior  along the time.

\subsection{Discriminating groups in the Chamber of Deputies}

In principle, a direct way to analyse the  dynamics of groups in the congress is by diving congressmen in their own political parties. For each party it is possible to quantify, for example, how aligned they are with the government and how faithful congressmen are to their own parties. These quantities can be evaluated for different roll calls and give us a notion of how groups behave over time. {In spite of} being a natural choice of groups, {this approach might be} problematic for some reasons. First because there are many  parties in the Chamber of Deputies, as shown in Table \ref{legislaturas_tab},  and they have different sizes and influence in the political scenario. Second, because parties break up and congressmen change parties along the same legislative periods~\cite{Vieira2019}. As a result, the fluctuation in these data are so important that one cannot extract a clear picture of the difference between stable and unstable regimes. Here also one might resort to the conclusions drawn from Fig. \ref{fig:votesinps} and realize that those issues dividing the congress are, as well those that divide the most the parties. To avoid these problems, in the following we use the sequence of agreements between the votes of the congressmen as entries in the $k$-means clustering algorithm to analyse how congressmen effectively aggregate. It allows us to identify some important groups in the congress and how {they} evolve in time.

\subsubsection{Identifying an effective {ruling coalition} and opposition {for} each presidential term}

We use the $k$-mean algorithm to divide the congressmen in groups. As explained in section~\ref{kmeans}, this type of algorithm  requires a pre-definition of the number $k$ of groups into which one wants to partition a given sequence of observations.
For each presidential term, we consider $k=2$ and divide all $\ndep$ congressmen in two groups, called $G_1^2$ and $G_2^2$. For each of these groups, we quantify their support to the president's party by using the quantity ${\cal S}_n^k$ defined in the section~\ref{Snk}.
Table \ref{tabelak2} shows the number of congressmen identified in each group, the proportion of these groups in terms of the total number of congressmen $\ndep$ and how each group aligns with the president's party.

\begin{table}
\caption{Absolute size  of the clusters $G_{i}^{2}$  in terms of number of deputies  (relative sizes  $G_{i}^{2}/\ndep$) and their support ${\cal S}_i^2$ to the president party in the correspondent period.}
\label{tabelak2}
\centering
\begin{tabular}{c|c|c|c|c}
 President & $G_1^2$ & $G_2^2$ & ${\cal S}_1^2$ & ${\cal S}_2^2$ \\ \hline \hline 
 Collor    &       356 (70\%) &       151 (30\%) &              0.84 &               0.4 \\

 Itamar    &      400 (79\%) &      106 (21\%) &              0.77 &              0.61 \\
\hline 
 FHC I      &      440 (77\%) &      133 (23\%) &              0.81 &              0.33 \\
\hline 
 FHC II      &      402 (71\%) &      161 (29\%) &              0.89 &               0.5 \\
\hline 
 Lula I     &      412 (76\%) &      133 (24\%) &              0.85 &              0.41 \\
\hline 
 Lula II     &      403 (75\%) &      133 (25\%) &              0.89 &              0.34 \\
\hline 
 Dilma I    &      438 (79\%) &      115 (21\%) &               0.8 &              0.45 \\
\hline 
 Dilma II    &      326 (63\%) &      195 (37\%) &              0.72 &               0.4 \\

 Temer     &      417 (78\%) &      120 (22\%) &              0.86 &              0.29 \\
\hline 
 Bolsonaro &      384 (74\%) &      133 (26\%) &              0.86 &              0.29 \\
\hline 
\end{tabular}
\end{table}

We  observe in that the group $G_1^2$ corresponds at least to $70\%$ of the total number of congressmen in all terms, except in the case of Dilma II. Moreover, the support quantity of this group ${\cal S}_1^2$ is around 0.8  in all terms, which  indicates that this group is aligned with the president's party. This suggests that this majority group can be associated to an {\it ruling coalition} with the president. The minority group $G_2^2$ has alignment with the president's party  ${\cal S}_2^2$ much  smaller than  ${\cal S}_1^2$. We associate this  minority group with an {\it effective opposition} to the government.

To check how much these effective ruling coalition $G_1^2$ and effective opposition $G_2^2$ coincide with an officially declared base and opposition to the government, we identified the legislative terms for which the election occurred in two rounds. 
When this is the case, in the second round, many parties organize in a coalition, declaring publicly support to one or the other candidate. The parties which belong to the coalition that won the election we identify as {\it self-declared ruling coalition} $B^{\rm sd}$, while parties in the coalition that lost the election are identified as {\it self-declared opposition}, $O^{\rm sd}$. 
Some parties do not declare any position before the election and were identified as \emph{non-declared}.  Tables~\ref{tab:2006} and \ref{tab:2014}  show, for  two legislative terms, the fraction of congressmen that belong to a party of the $B^{\rm sd}$ that is identified as well as either an effective basis ($G_1^2/B^{\rm sd}$) or effective opposition, ($G_2^2/B^{\rm sd}$), and the same for the self declared opposition ($G_i^2/O^{\rm sd}$)  and non-declared position. Observing the diagonal of these tables one can see that $G_1^2$ corresponds to $98.7\%$ for Lula II and $89\%$ for Dilma II  of the self-declared basis $B^{\rm sd}$.  $G_2^2$ coincides with $84.7\%$ in the case of Lula II and with $75\%$ for Dilma II of the self-declared opposition. These high values occur for all the legislative terms for which we could do this analyses and they justify our interpretation of the groups encountered by the $k$-means algorithm as ruling coalition and opposition.


\begin{table}[h]
\caption{Lula II - Crosstabulation of identified clusters $G_i^2$ and self declared party affiliation at the second round of the election. Footnotes in these tables specify the parties that publicly declared support for one candidate or the other.}
\label{tab:2006}
\begin{tabular}{ r|c|c||c}
\multicolumn{1}{r}{}
 &  $G_1^{2}$ & $G_2^{2}$ & Total \\
\cline{1-4}
$B^{\rm sd}$ \footnote{
PCdoB, 
PP,   
PR,   
PRB,  
PSB,  
PT, 
PTB,  
}    & 232 (98.7\%) & 3 (1.3\%) & 235\\
\cline{1-4}
$O^{\rm sd}$ \footnote{         DEM (PFL), 
         PPS,   
         PSDB   
         } 
& 22 (15.3\%) & 122 (84.7\%) & 144\\
\cline{1-4}
Non Declared \footnote{
         PAN,   
         PHS,   
         PMN,   
         PSOL,  
         PTC,   
         PTdoB, 
         PV,    
         PDT,   
         PMDB,  
         PSC   
         }
& 149 (94.9\%) & 8 (5.1\%) & 157\\ \hline \hline
Total             & 403 (75.2\%) & 133 (24.8\%) & 536\\ 
\end{tabular}
\end{table}

\begin{table}[h]
\caption{Dilma II - Crosstabulation of identified clusters $G_i^2$ and self declared party affiliation at the second round of the election. Footnotes in these tables specify the parties that publicly declared support for one candidate or the other.} 
\label{tab:2014}
\begin{tabular}{ r|c|c||c }
\multicolumn{1}{r}{}
 &  $G_1^{2}$ & $G_2^{2}$ & Total \\
\cline{1-4}
$B^{\rm sd}$ \footnote{
        PT, 
        PMDB,
        PSD, 
        PP, 
        PR, 
        PROS, 
        PDT, 
        PCdoB, 
        PRB 
} 
&  275 (89.3\%) & 33 (10.7\%) & 308\\
\cline{1-4}
$O^{\rm sd}$ \footnote{
        PSDB, 
        PMN, 
        SD, 
        DEM,
        PEN,
        PTN, 
        PTB, 
        PTC, 
        PTdoB,
        PSB, 
        PV, 
        PSC,
        PPS,
        PSDC,
        PHS, 
        PSL, 
        PRP, 
        PRTB 
} 
& 51 (24.5\%) & 157 (75.5\%) & 208\\
\cline{1-4}
Non Declared    \footnote{PSOL
} 
  & 0 (0\%) & 5 (100\%)  & 5\\ \hline \hline
Total              & 326 (62.6\%) & 195 (37.4\%)  & 521\\
\end{tabular}
\end{table}

\subsubsection{Interpretation of the agreement between congressmen}

\begin{figure*}
\centering
\includegraphics[width=0.4\columnwidth]{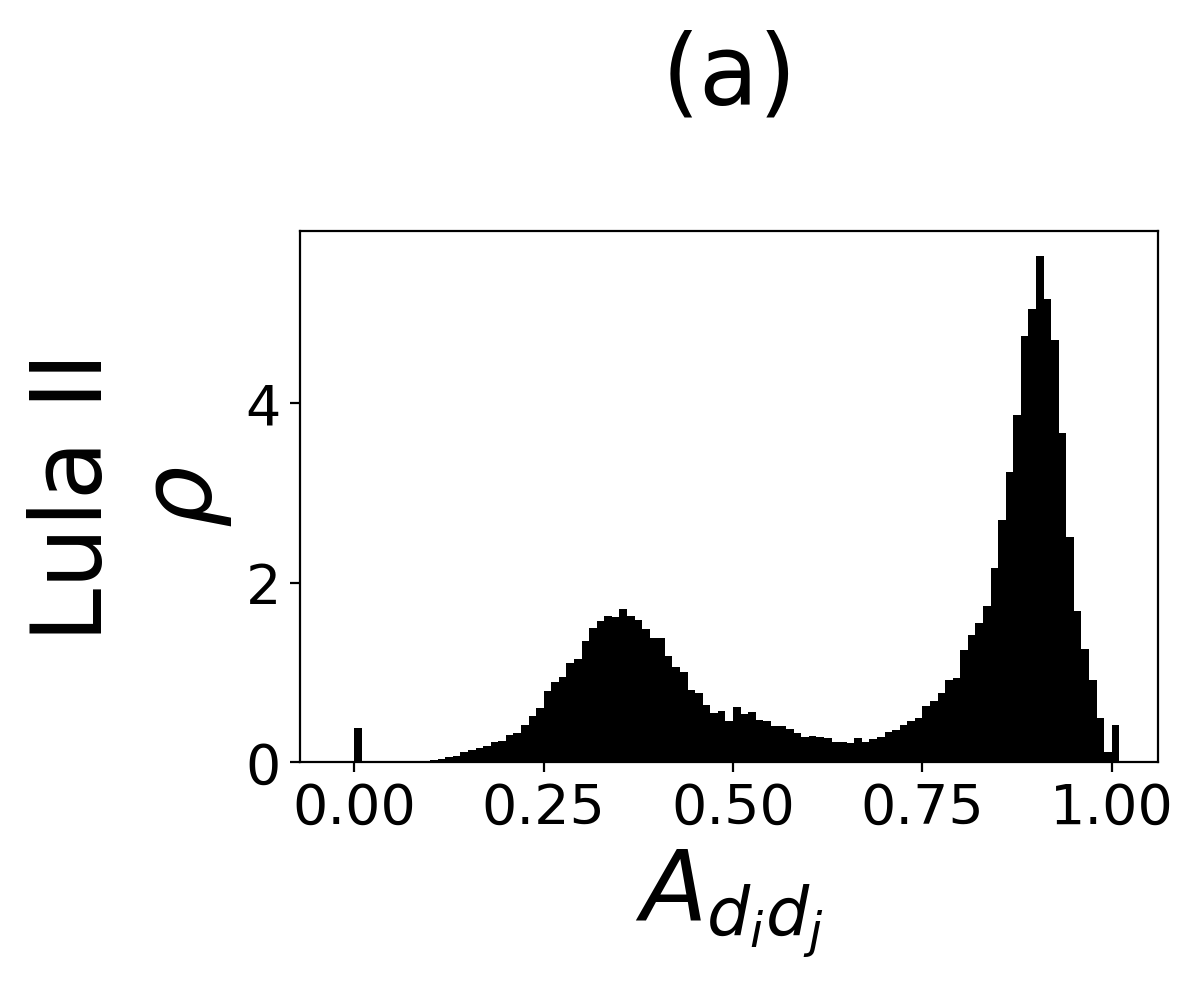}
\includegraphics[width=0.4\columnwidth]{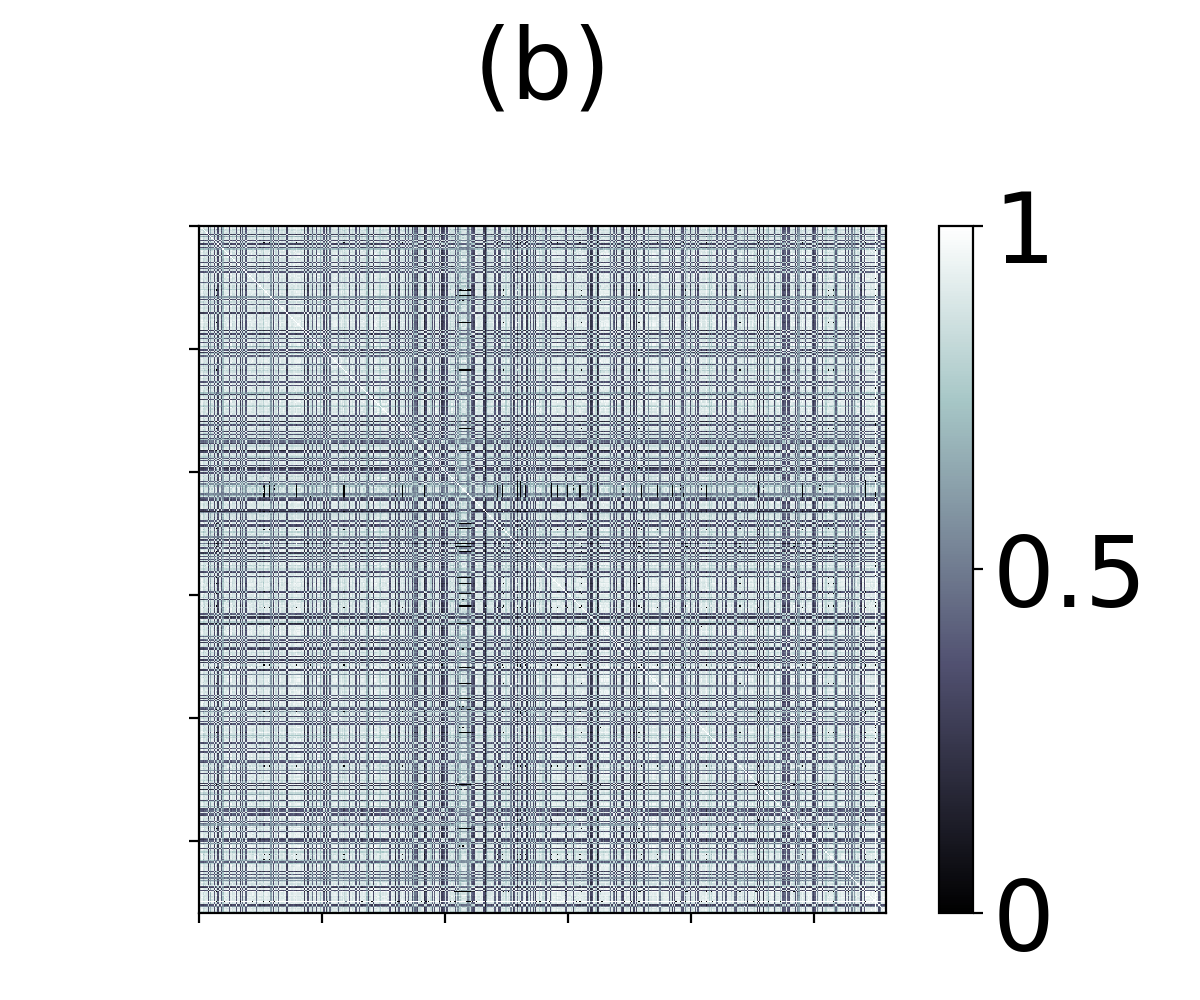} 
\includegraphics[width=0.4\columnwidth]{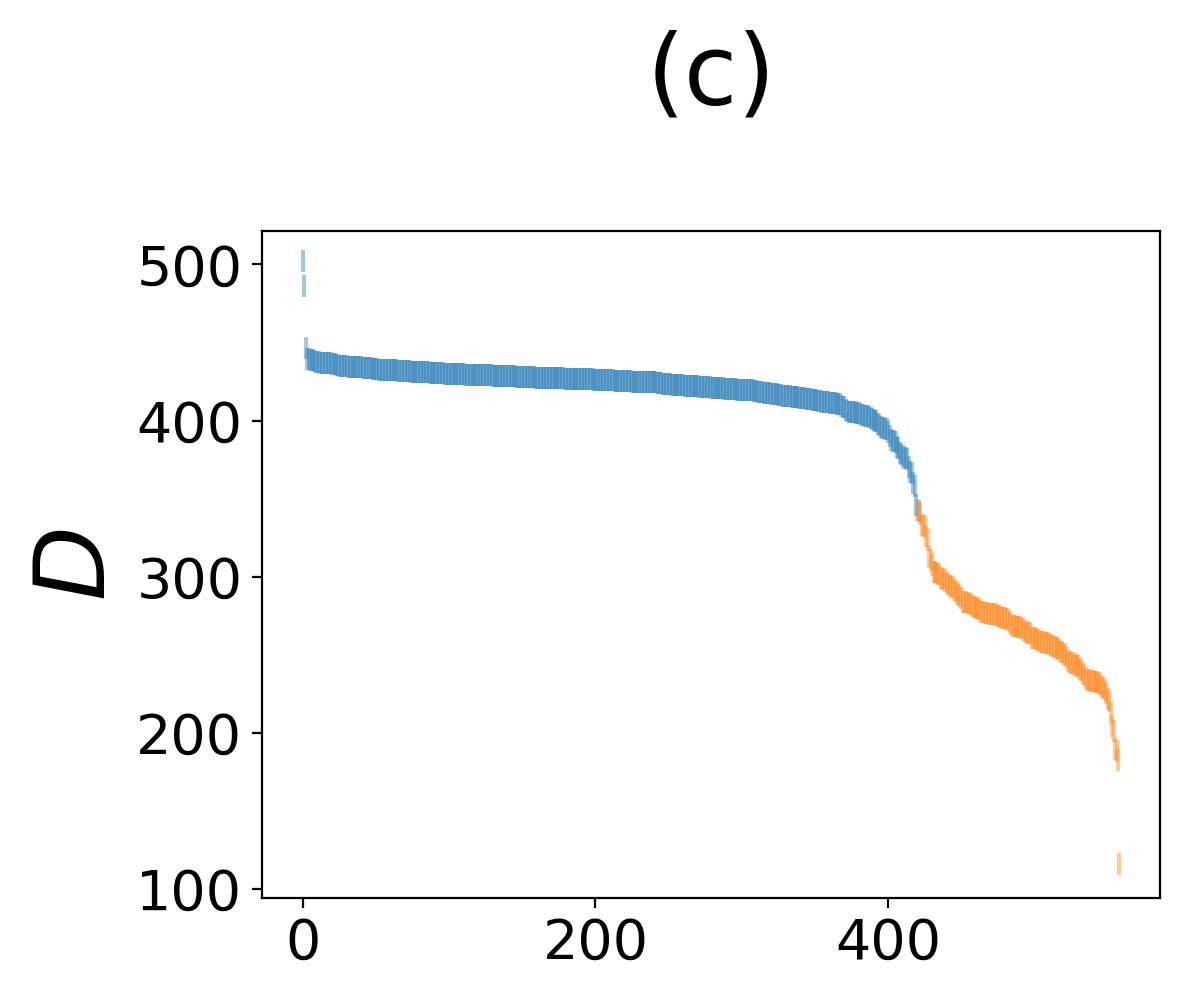}
\includegraphics[width=0.4\columnwidth]{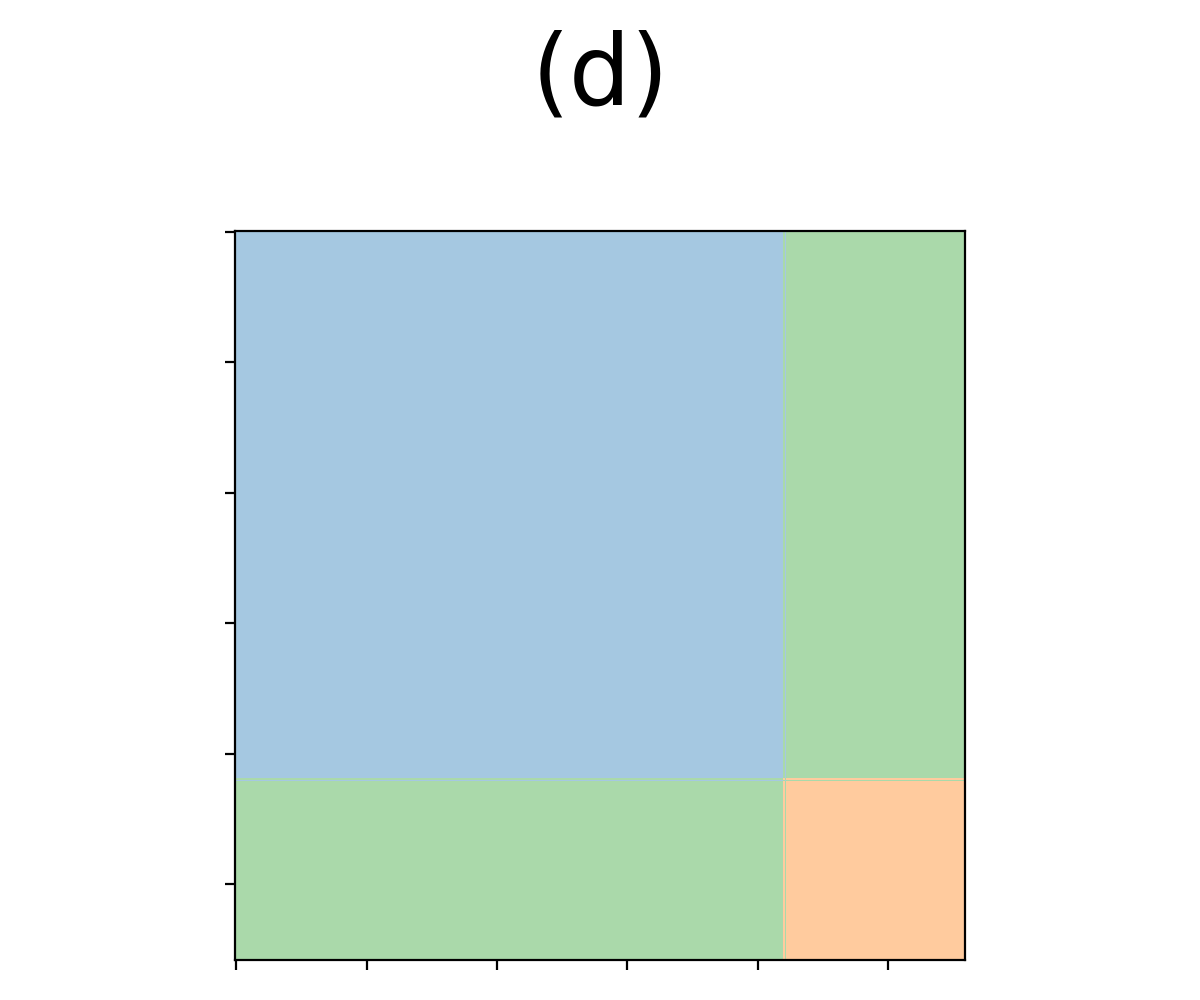}
\includegraphics[width=0.4\columnwidth]{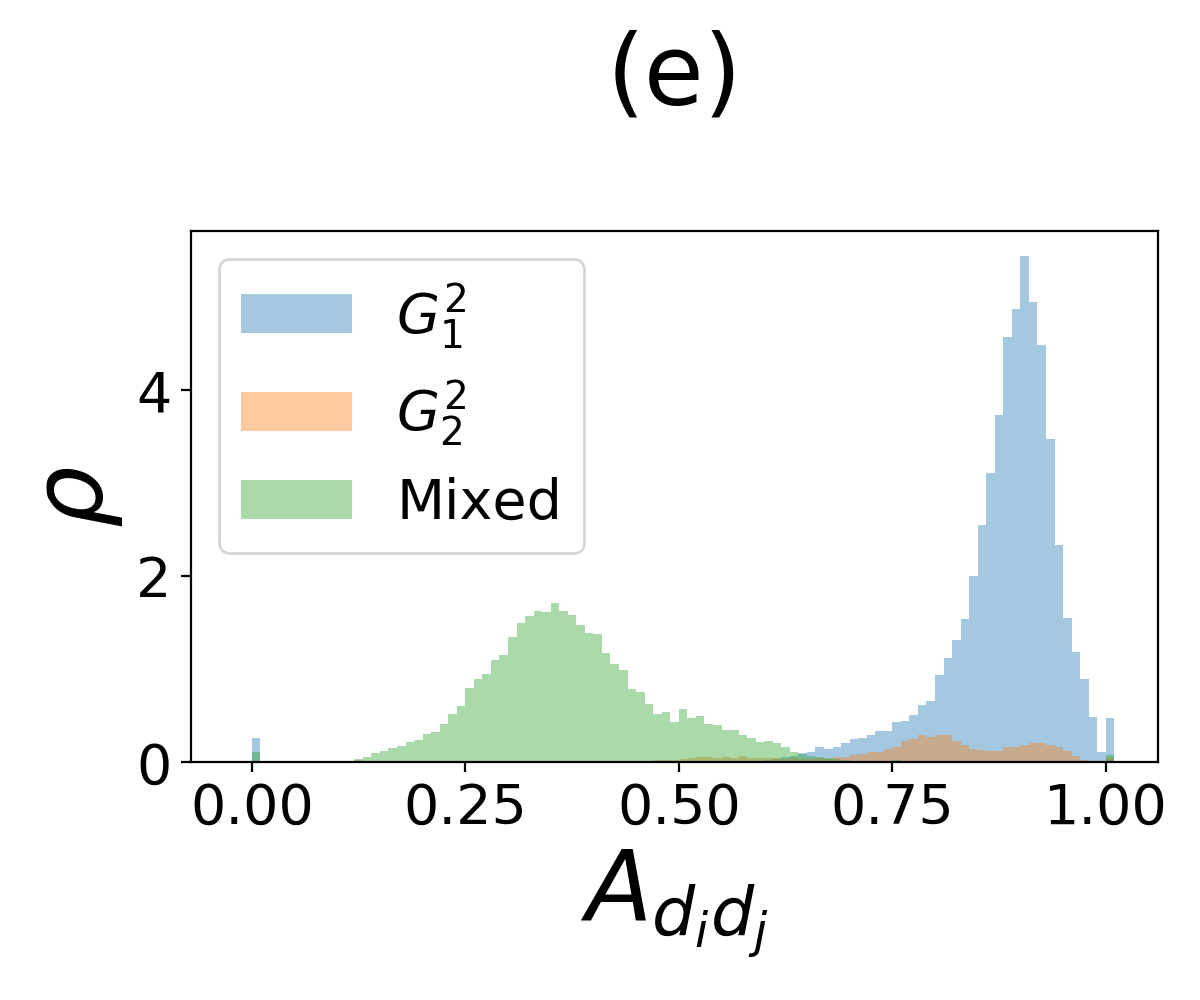}
\includegraphics[width=0.4\columnwidth]{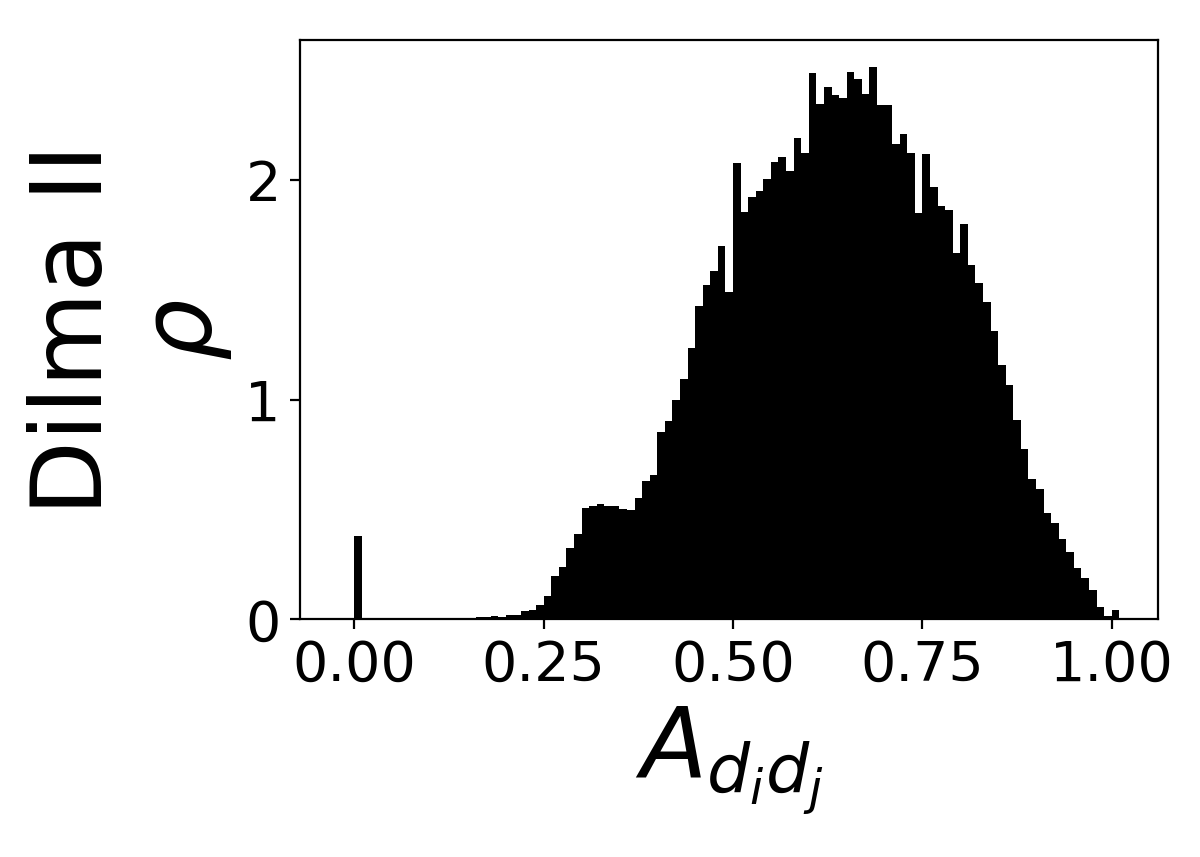}
\includegraphics[width=0.4\columnwidth]{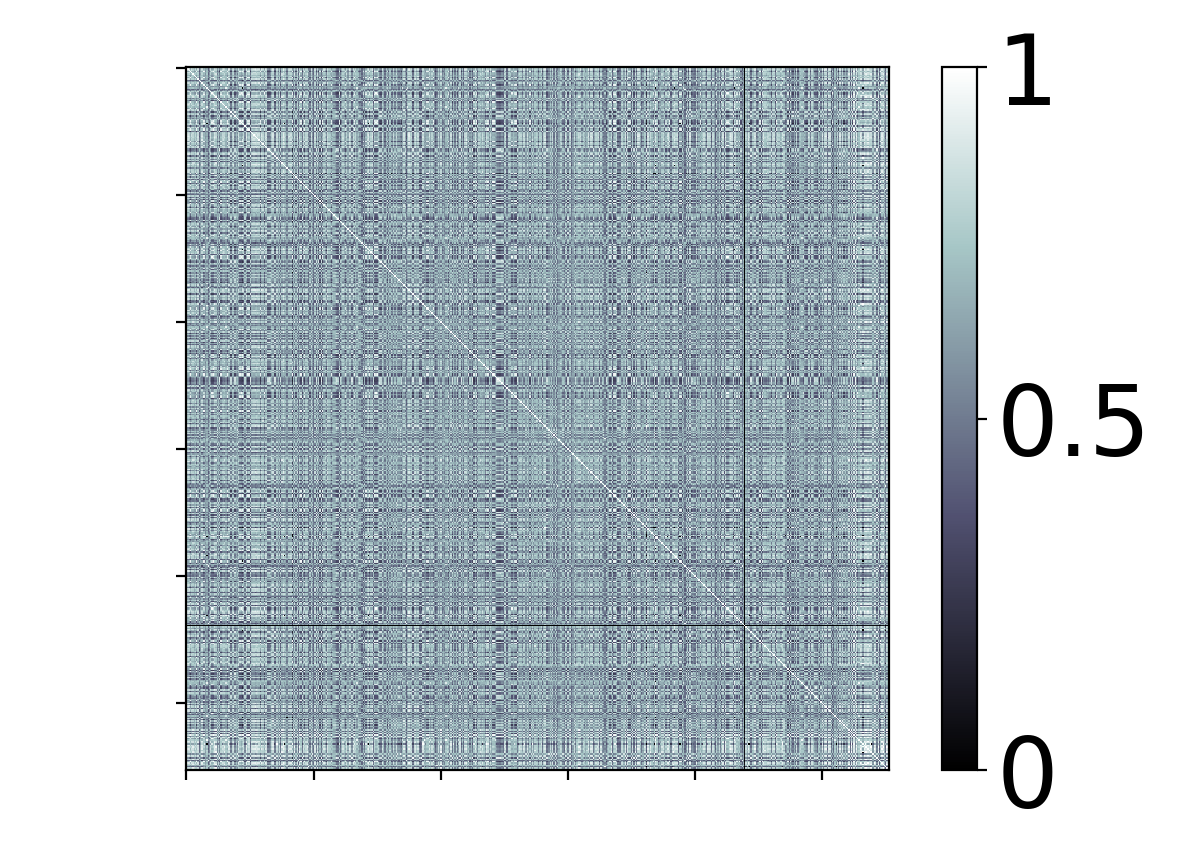} 
\includegraphics[width=0.4\columnwidth]{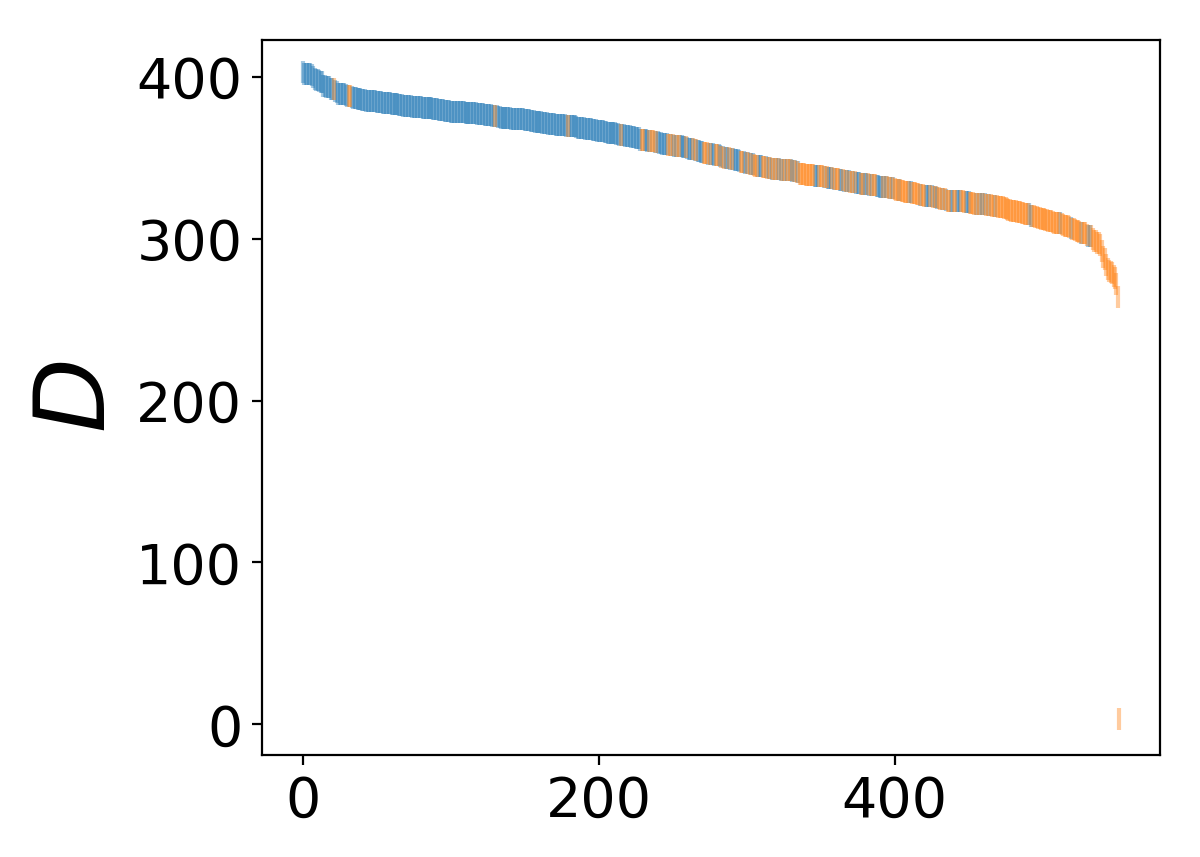}
\includegraphics[width=0.4\columnwidth]{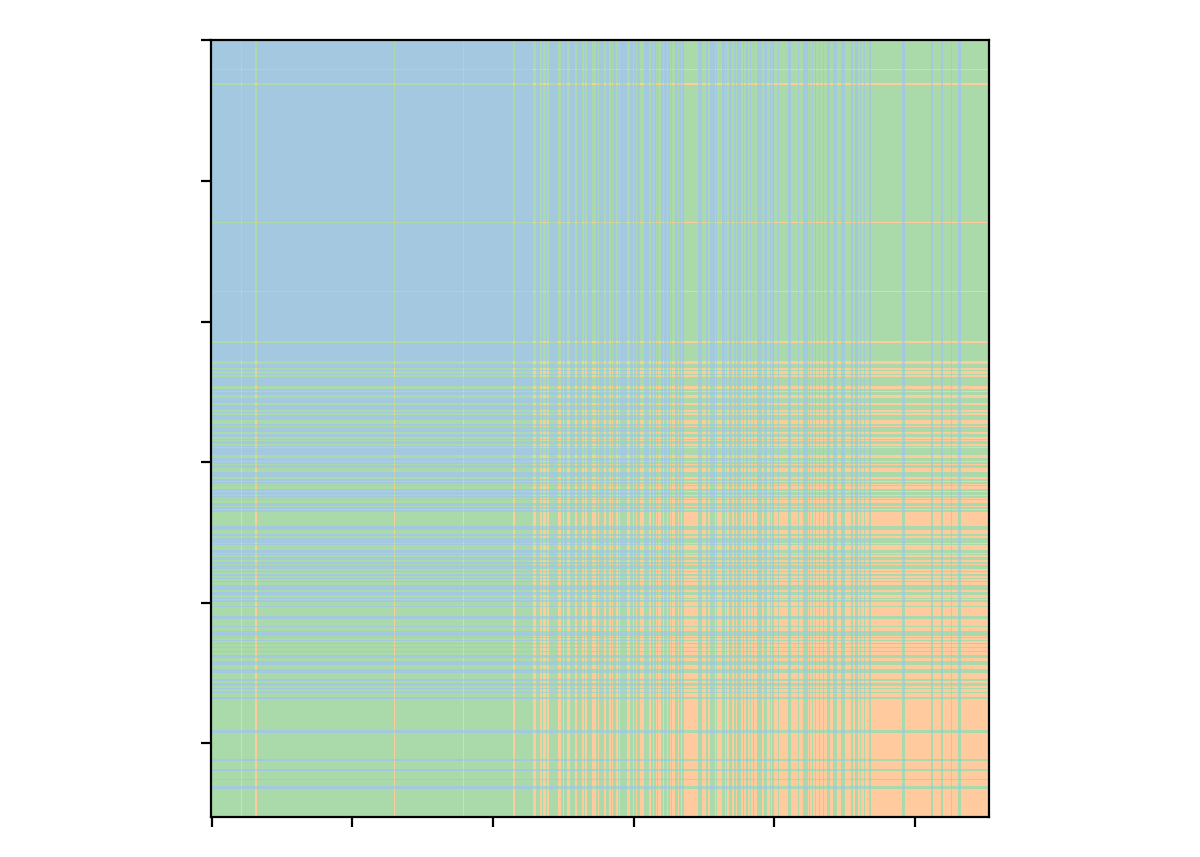}
\includegraphics[width=0.4\columnwidth]{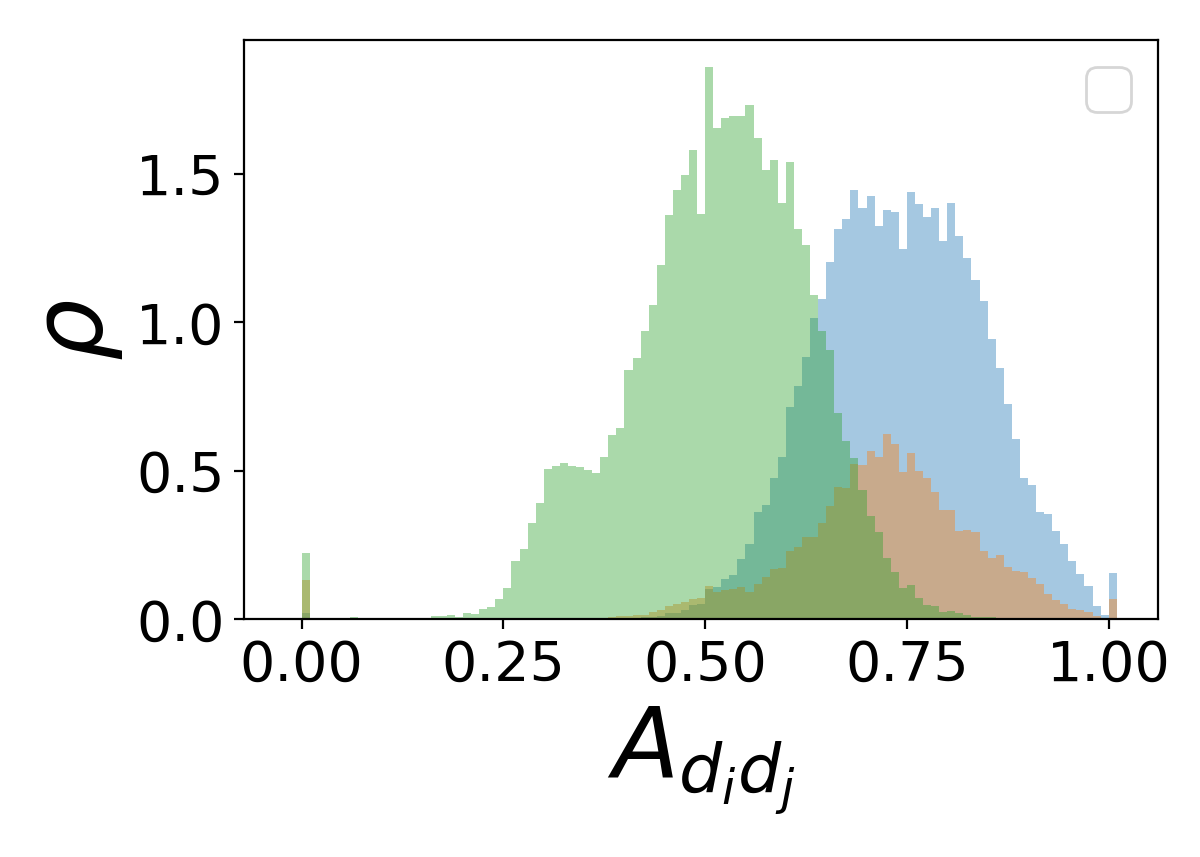}
\includegraphics[width=0.4\columnwidth]{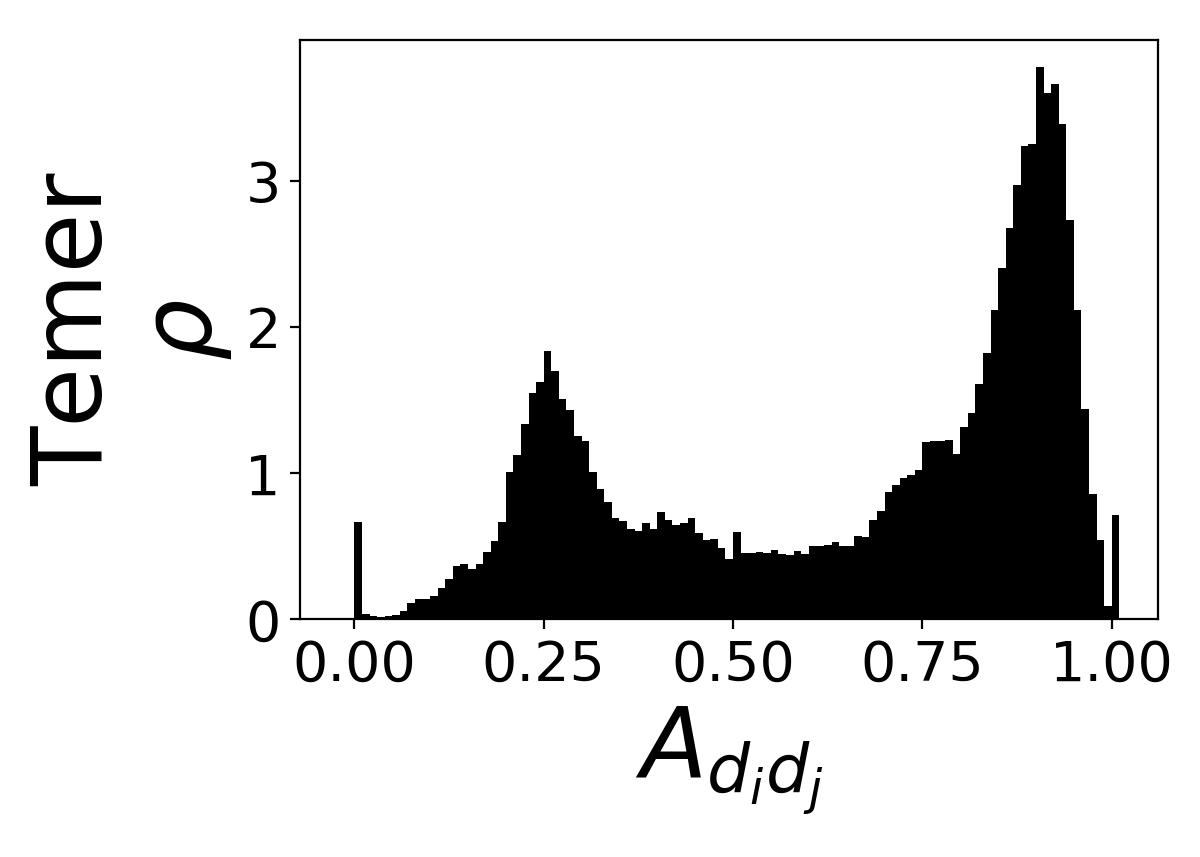}
\includegraphics[width=0.4\columnwidth]{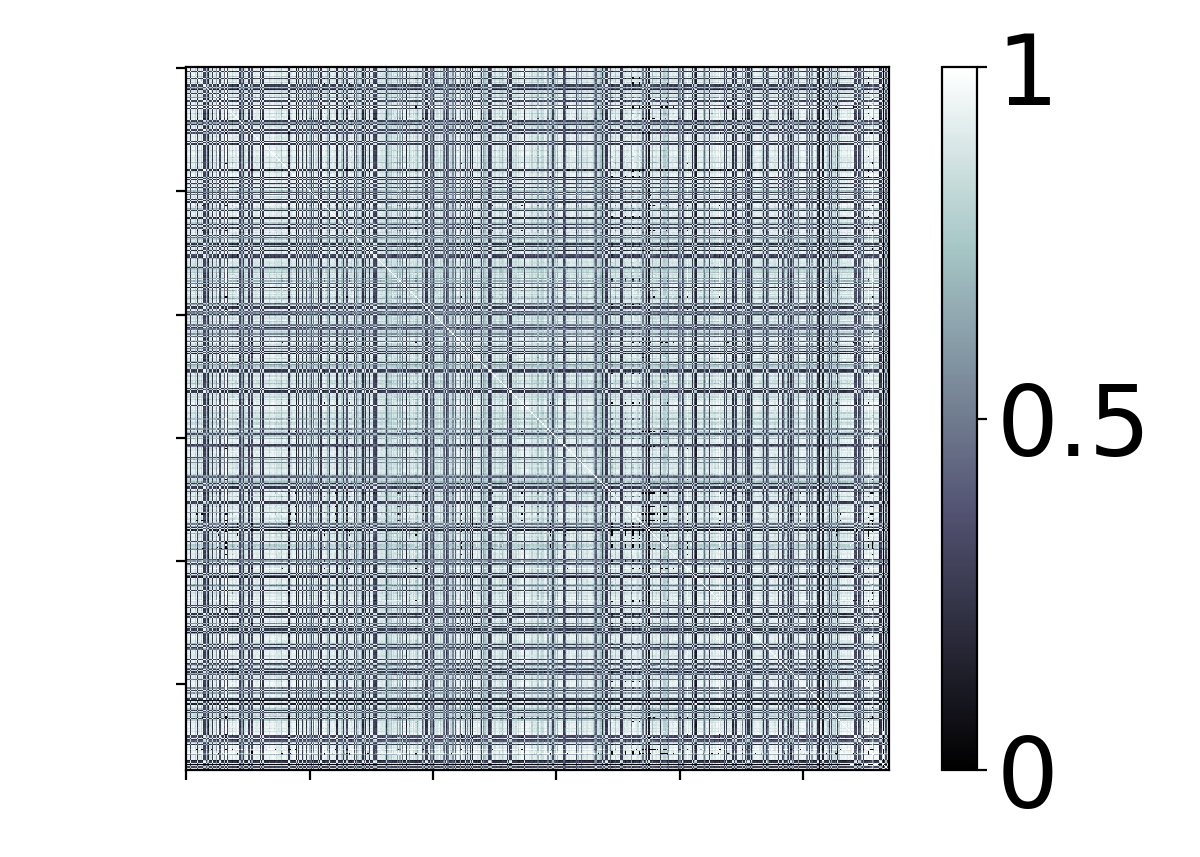} 
\includegraphics[width=0.4\columnwidth]{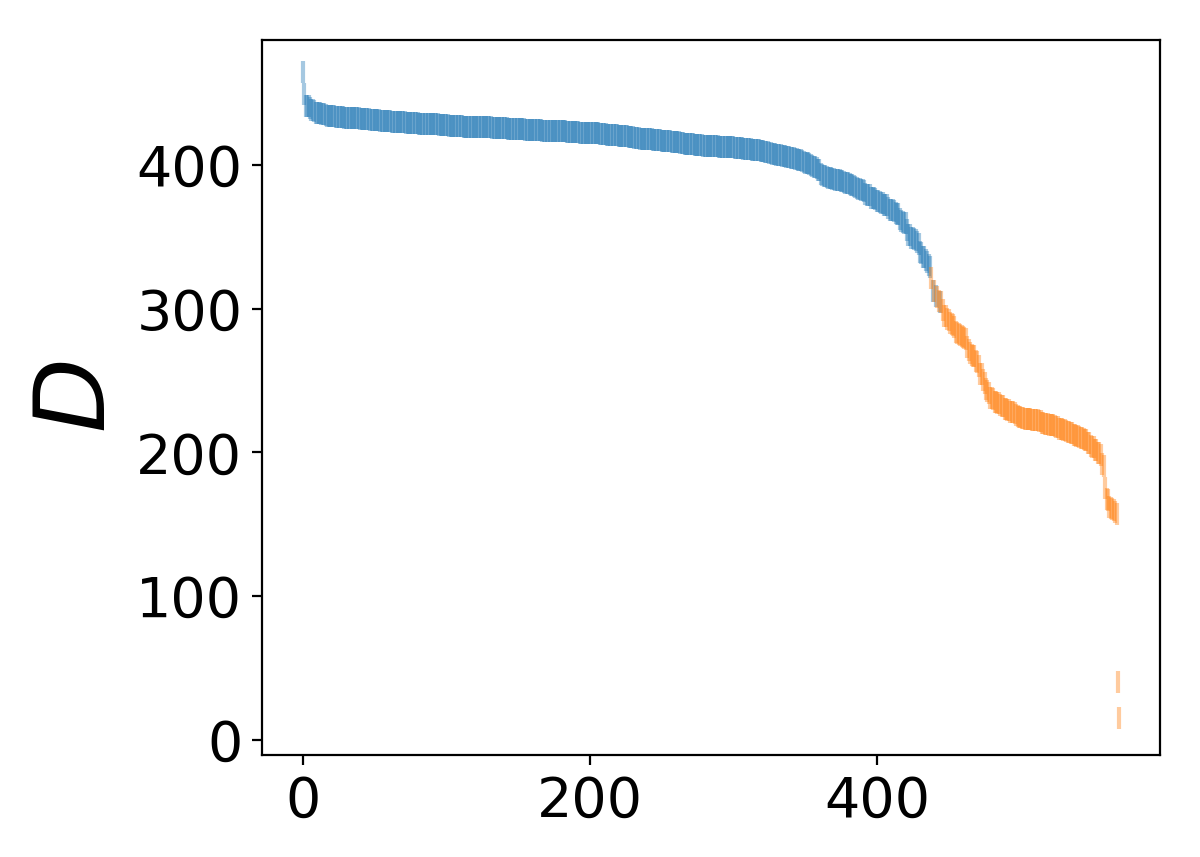}
\includegraphics[width=0.4\columnwidth]{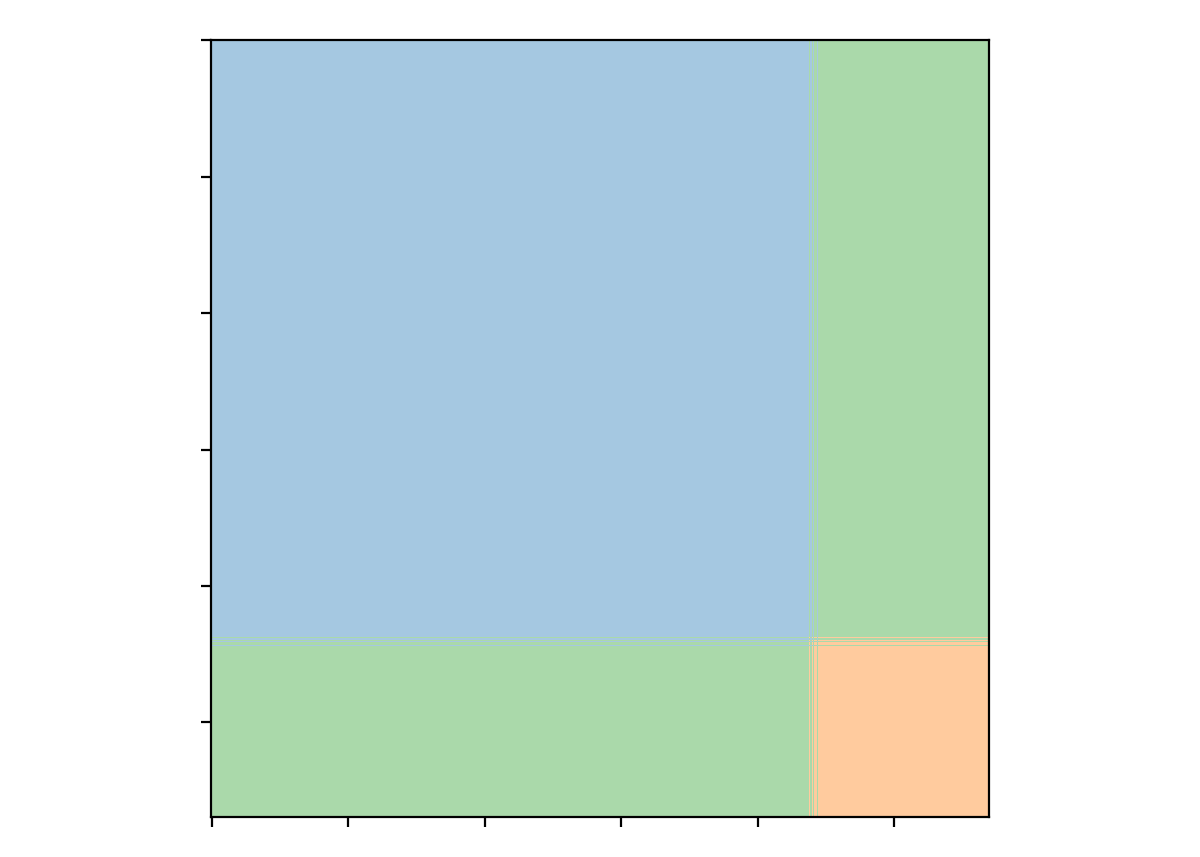}
\includegraphics[width=0.4\columnwidth]{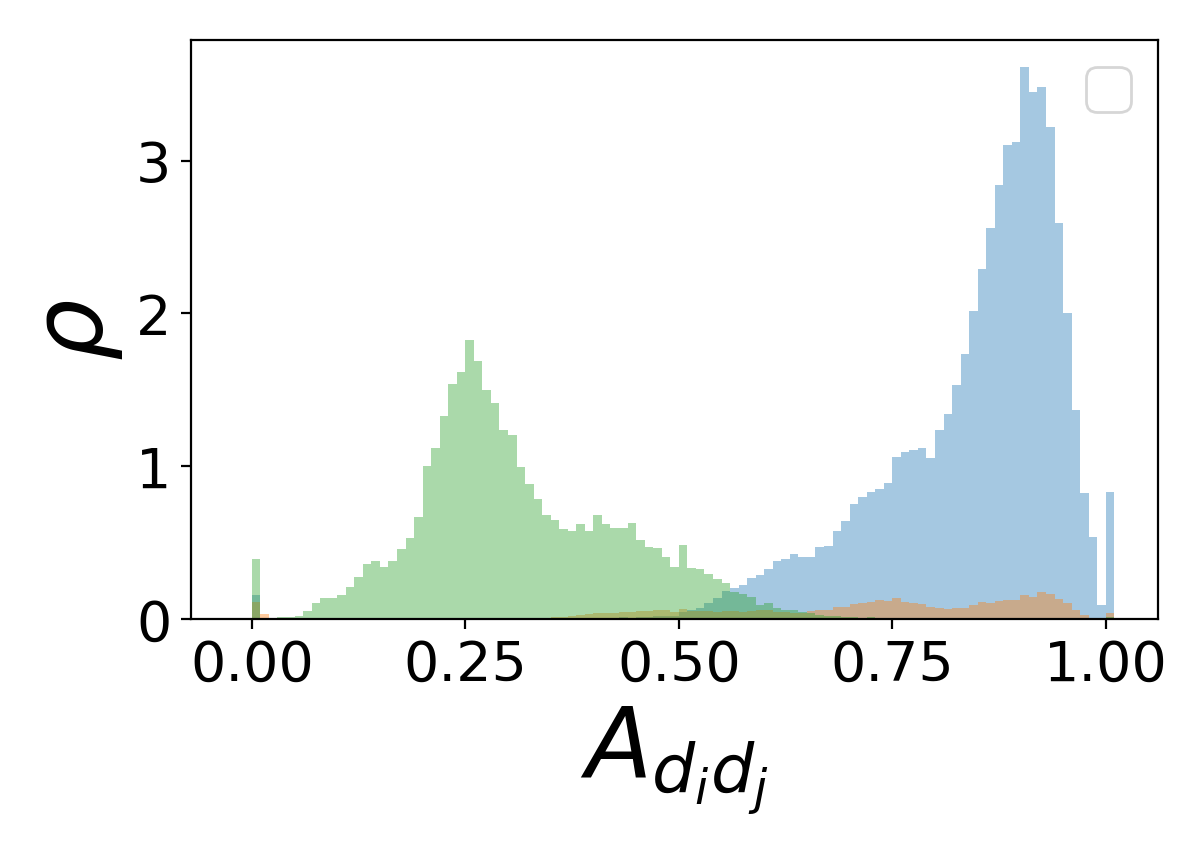}
\caption{For the periods  LulaII, Dilma II and Temer, we show (a) the agreement matrix distribution, (b) the agreement matrix with the agreement according to the color-bar to the right, (c) ranked degrees centrality ${\cal D}$ of the congressmen colored by which cluster they were assigned to (blue for $G_1^2$ and orange for $G_2^2$), (d) ordered agreement matrix with congressmen ranked by degree centrality  and colored blue if both belong to $G_1^2$, orange if both belong to $G_2^2$ and green if one belongs to $G_1^2$ and the other to $G_2^2$, and (e) distribution of agreement matrix separated by the same coloring criterion of (d).}
\label{fig:Aplicak2}
\end{figure*}

In this subsection we use the separation in two groups discussed in the previous subsection to interpret the peaks in the distribution of agreements and how good this  separation is to identify different behaviors in the Congress. Fig.\ref{fig:Aplicak2} shows the data for two different legislative periods: Lula II that is typical of a stable period and the Dilma II and Temer,  the two parts of a term where there was an impeachment.

The first line of Fig.\ref{fig:Aplicak2}-a shows $\rho(\Aij)$ for the Legislative period correspondent to Lula II, which has been already presented in Fig.\ref{fig:distribuicoesAij} and is repeated here to be easily comparable with the Dilma II and Temer periods, shown in the second and third lines of Fig.\ref{fig:Aplicak2}-a, respectively. We remind that Dilma II is defined as the period before the beginning of impeachment process in the Legislative period called Dilma-Temer. The exact dates used to separate the two periods are shown in Table \ref{legislaturas_tab}. Fig.\ref{fig:Aplicak2}-b show the matrix of entries $\Aij$ as a function of the indexes of the deputy $\depi$ and $\depj$. We then measure the degree centrality of the $\depi^{th }$ congressman, defined as $\cal{D}_{\depi} = \sum_{\depj} \Aij$ and introduce a label ${\rm m}$ to rank the congressmen in descending order of ${\cal D}_{\rm m}$,  ${\cal D}_1>{\cal D}_2>...{\cal D}_{\ndep}$. This ordered quantity  $\cal{D}_{\rm m}$ as a function of ${\rm m}$ is shown in Fig.\ref{fig:Aplicak2}-c (we omit the index ${\rm m}$ for the sake of cleanness of the figure). The blue colors shown in this figure represents the congressmen of the group $G_1^2$ and orange represents the congressmen in $G_2^2$. We use this rank defined by  $D_{\rm m}$ to reorder the matrix of agreements presented in Fig.\ref{fig:Aplicak2}-d. The colors in this reordered matrix are defined using the following criterion: if two congressmen $\depi, \depj$ belong to the same cluster $G_1^2$, $\Aij$ is colored blue; if both congressmen belong to cluster $G_2^2$, they are colored orange; and  if each one  belongs to a different group, they are colored green and referred to as ``mixed term''. This same notation of colors is used to represent the distribution of $\Aij$ for these tree cases in Fig.\ref{fig:Aplicak2}-e. In this case the distributions are normalized in such a way that {the sum of the areas  under the curves of the distribution is one}, observing the proportion between the size of the groups. Our sorting strategy could be contrasted with the one adopted in \cite{MarencoPlosOne2020} which is to use the Prim algorithm to determine the minimum spanning tree for their set of distances (agreements). This strategy considers only the strongest agreements between congressmen, while ours is taking into account all information (all agreements).

We can now critically analyse Fig.\ref{fig:Aplicak2} to observe differences between these legislative terms. The first observation was already anticipated in section \ref{sec:distAij}: the period called Dilma II has a unimodal distribution, while Temer recovers the two peaks observed in Lula II and other stable political periods. Without any order,  matrices of agreement shown in the Fig.\ref{fig:Aplicak2}-b do not allow us to distinguish any difference between these periods. In Figs.\ref{fig:Aplicak2}-c, we observe that in Lula II and Temer there is a step in the quantity  ${\cal D}_{\rm m}$, and the two levels of the step correspond to the two groups classified by the clustering algorithm. On the other hand, in Dilma II there is a continuous decay of  ${\cal D}_{\rm m}$. Ordered matrices represented in Fig.\ref{fig:Aplicak2}-d  show a very well defined division in two groups during the stable political periods as Lula II and Temer, but in Dilma II, where there was an impeachment, the division in two groups is not enough to capture the complexity of the groups' organization in the congress.  Fig.\ref{fig:Aplicak2}-e show that during Lula II and Temer, the blue color, that correspond to the effective ruling coalition, is large, while the orange group, which was shown to be an effective opposition, is very small. The green group, which are the mixed terms, has a much smaller size in the periods Lula II and Temer, while is the most important  distribution in Dilma II. This mixed group has a peak at smaller values of $\Aij$ and this is just saying that congressmen from two different groups have smaller agreement in their sequence of votes, while for the effective ruling coalition and opposition the average agreement is high inside of each group, which indicates strong cohesion among them.

The discussion above for the two typical legislative terms can be extrapolated to the other six terms analyzed in this work with the same observations.  We present a schema of these characteristics for all legislative terms in the Fig.\ref{fig:resumo}.

To summarize, our analyzes have shown that, during stable political periods, the government {forms an effective ruling coalition}  which contains at least 70\% of the congressmen of the congress and this group presents a strong support to the president's party. In legislative terms characterized by an impeachment, the separation into two effective groups is not enough to discriminate the matrix of agreement {between congressmen} completely. In the next section we propose to analyse how these two groups evolve along time in the different legislative terms and also a further analysis in terms of considering more than two  groups to capture what happens in unstable periods.


\subsection{Time evolution of the groups in the Chamber of Deputies}
\label{sec:timeevolution}

\begin{figure}[h!]
\centering
\includegraphics[width=0.49\columnwidth]{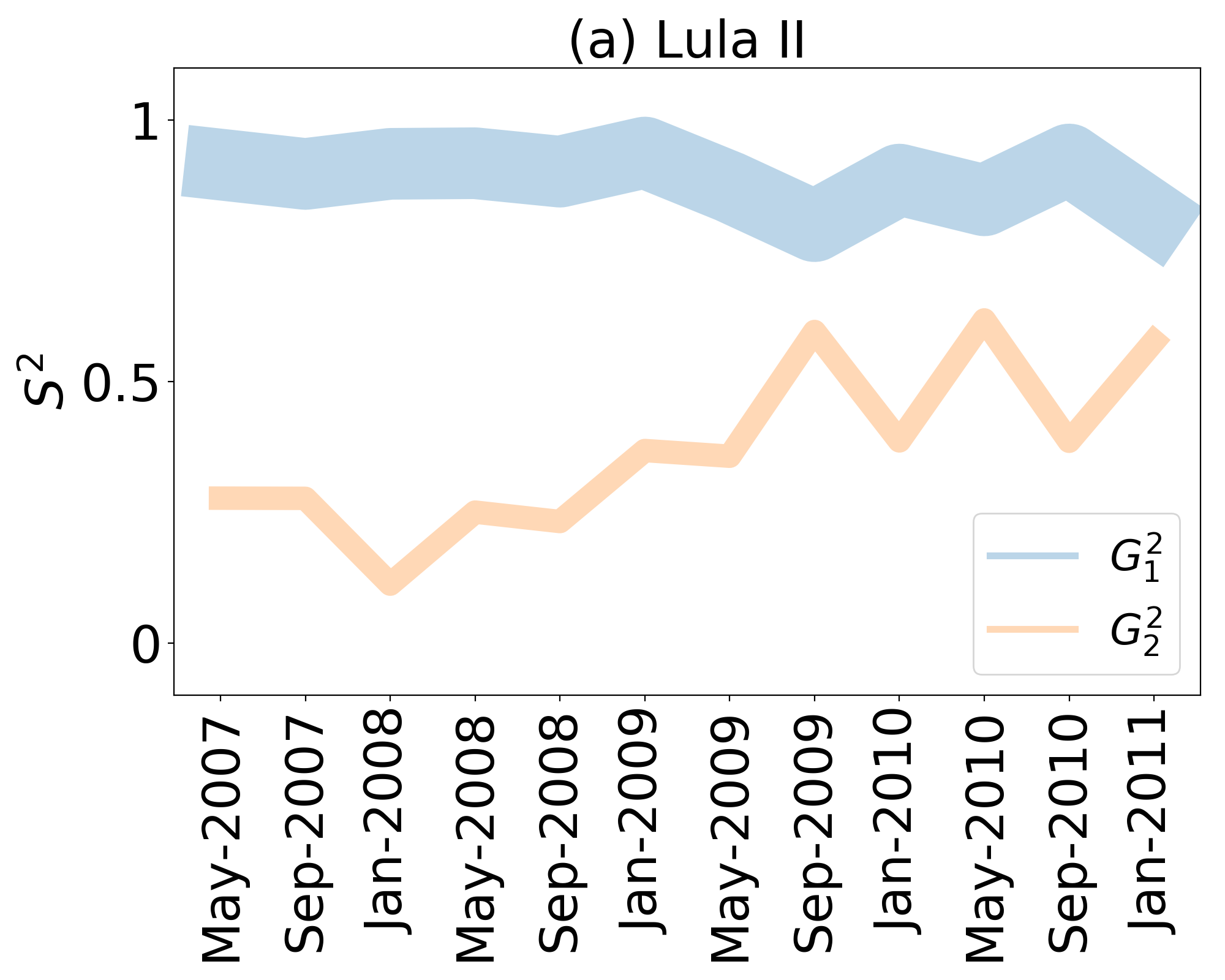}
\includegraphics[width=0.49\columnwidth]{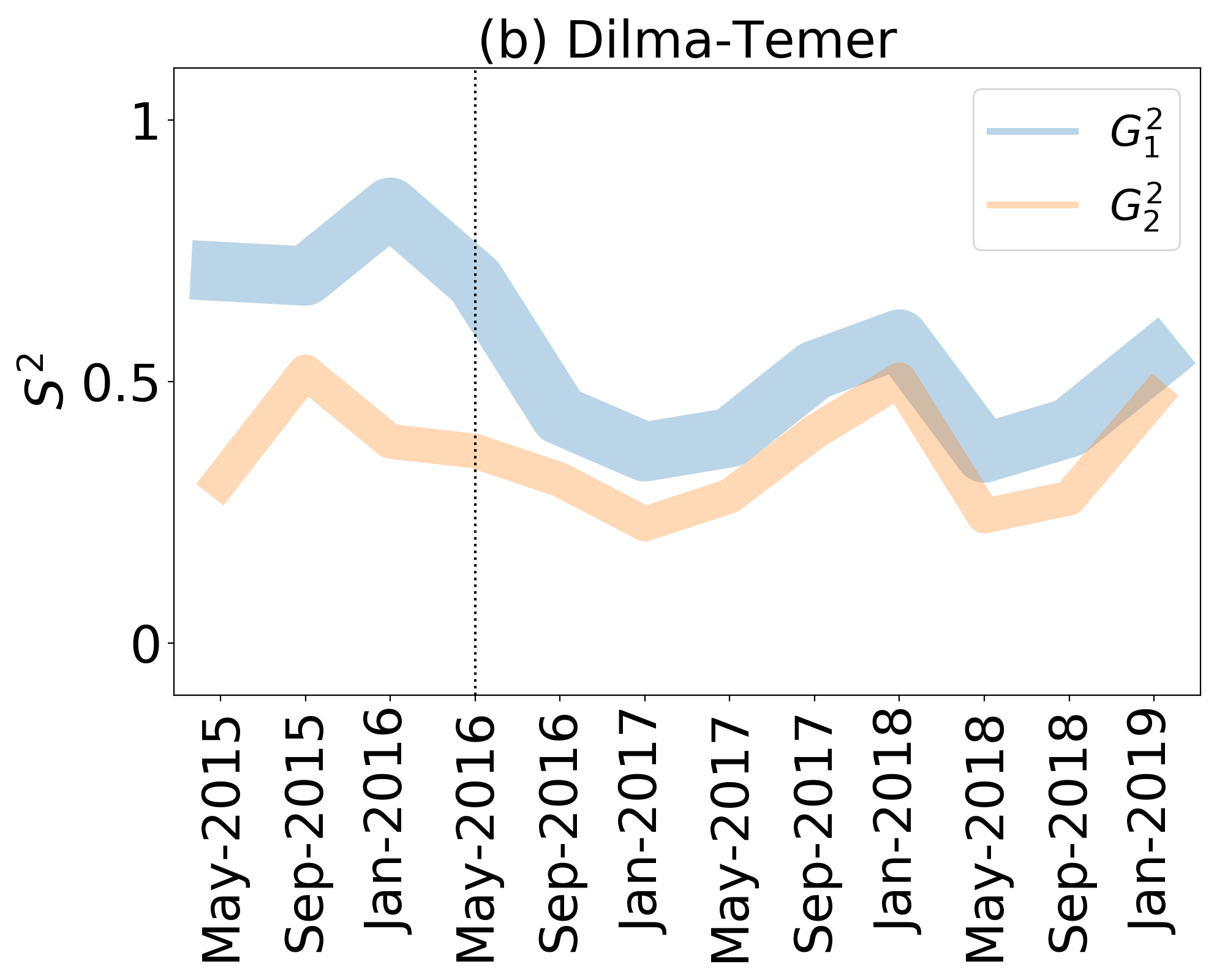}
\includegraphics[width=0.49\columnwidth]{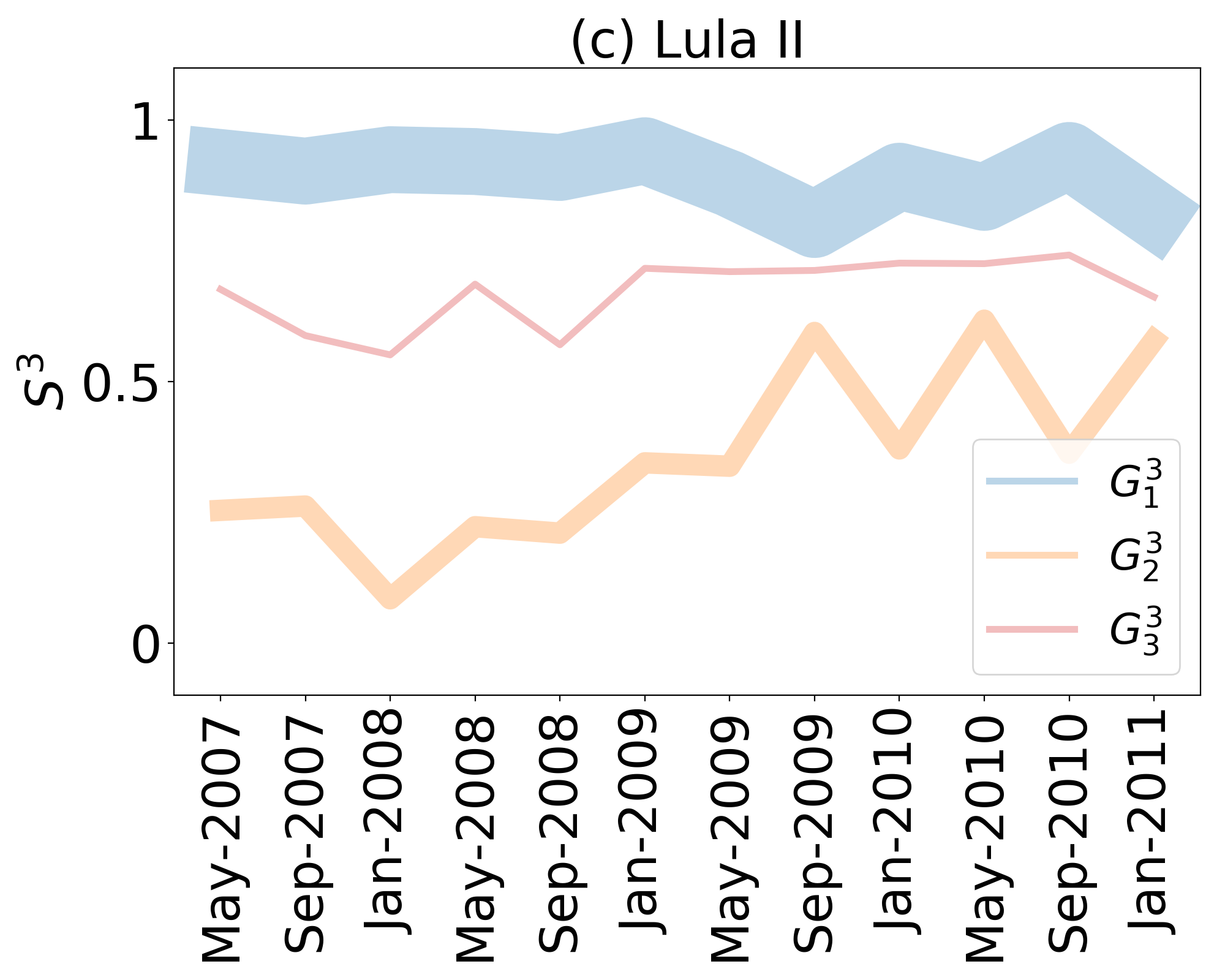}
\includegraphics[width=0.49\columnwidth]{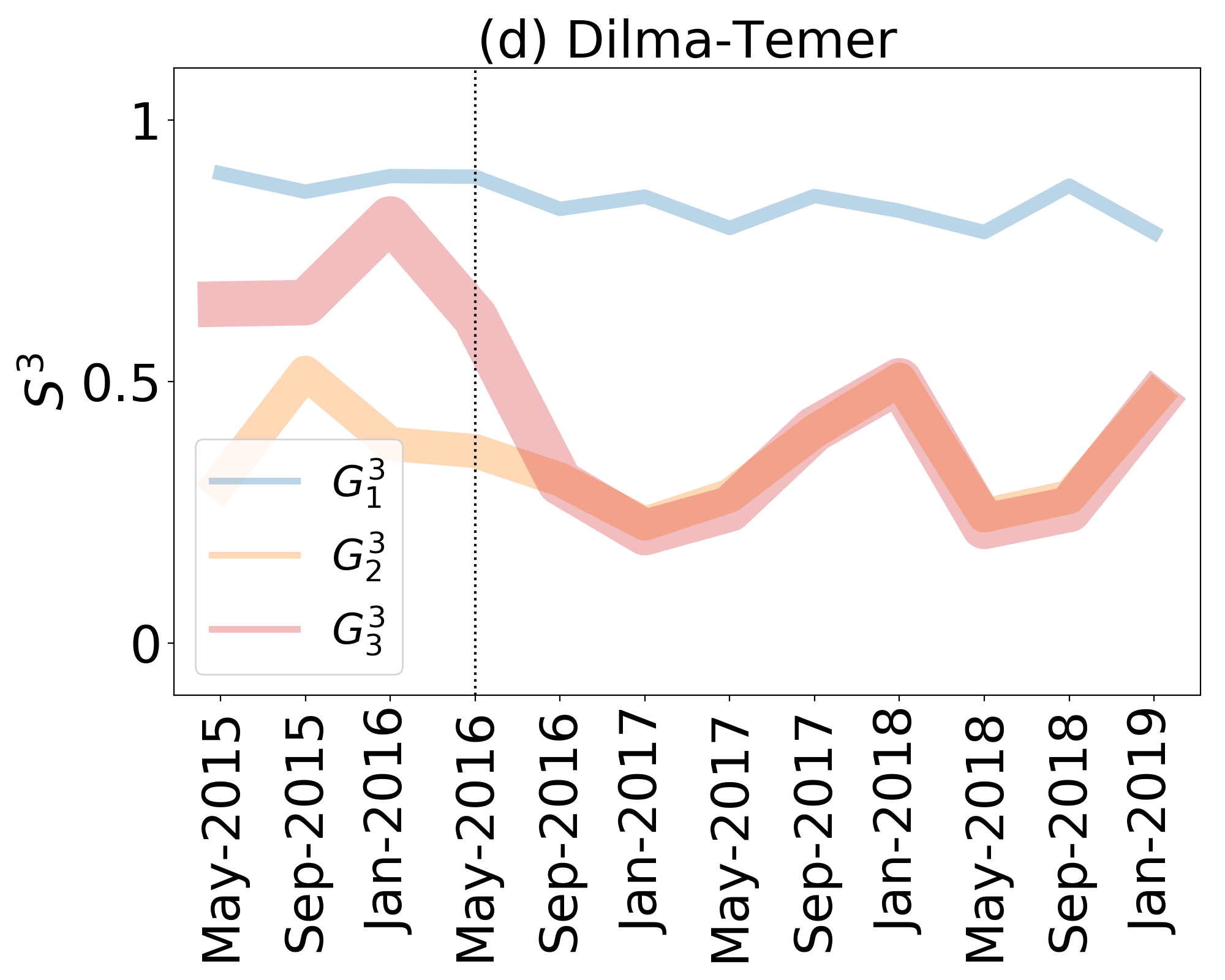}
\caption{Evolution of clusters support to the president's party. The clusters were found with k-means and using the whole Lula II period and from the beginning of Dilma II period up to the beginning of her impeachment process, indicated by the vertical line. The width of the line is proportional to the cluster size.  The legend indicates the group name.}
\label{fig:kp}
\end{figure}

An average measure over the whole legislative terms might not allow one to understand what happens before an impeachment or how this is compared to a  politically stable legislative term. To verify how the support to the government of different groups behaves along time within a legislative term, we  use the following procedure:  We first cluster groups of congressmen using $k$-means.
Once we have these groups defined, we separate the roll calls in windows of four months and, for each period, we measure the support of each group to the president's party, $S_n^k$, as defined in Eq. (\ref{kappa}). The result of this measure in time is shown in Fig.\ref{fig:kp} for the same two legislative terms previously discussed, Lula II and Dilma-Temer.
Figures on the top show the case where the congressmen are segregate in two effective groups. We emphasize that, for the Lula II term, groups were defined using the whole presidential term, while for the term Dilma-Temer, groups were defined using the period before the impeachment (the exact dates are shown in Table~\ref{legislaturas_tab}). Also note that, that Figs.\ref{fig:kp}.-b, d show the whole term Dilma-Temer, but what is represented is the alignment with the Dilma Rousseff's party. In these figures, the line width of each group is proportional to their size and the the notation of colors is the same as in the previous section: blue denotes effective {ruling coalition}, $G_1^2$, and orange the effective opposition, $G_2^2$. Vertical lines in Fig.\ref{fig:kp}-b, d indicates the end of Dilma Rousseff's  mandate ({i.e. the moment Dilma was removed from government because the impeachment process begun.}).

One can observe that, during the whole period of Lula II, the effective {ruling coalition} keeps its support to the president's party and that the effective opposition has a smaller support value that increases at the end of his mandate. This behavior is observed in other politically stable legislative terms and is in line with the observation in a recent work that the  polarization decreases as the end of mandate is approached \cite{MarencoPlosOne2020}. In the Dilma-Temer period, the situation is very different: i) the support and the size of the effective {ruling coalition} is smaller than in the Lula II period and ii) the group associated to an effective {ruling coalition} in the Dilma's government decreases its support $S_1^2$ to a value comparable to the one of the effective opposition $S_2^2$ after the impeachment.

To better understand this behavior we repeat the previous analysis, but this time using $k$-means with $k=3$ to separate the congressmen in clusters  $G_1^3$, $G_2^3$ and $G_3^3$. For each of these groups, we measure their support to the president's party as a function of time and show the results in Figs.\ref{fig:kp}-c,d. In these figures, {red}  is the third group identified in this process. As previously, the line width is proportional to the group sizes. In the period Lula II, we observe that the third group is very small and  basically is aligned to the president's party as the effective {ruling coalition}. However, in {Dilma II}, the third group  $G_3^3$ is significantly bigger than the effective {ruling coalition} $G_1^3$. Moreover, its support to Dilma Rousseff's measured by $S^3_3$ clearly decays and joins the  effective opposition {$G^3_2$}. The effective {ruling coalition}, shown in blue $G^3_1$, is really small {but} keeps aligned during the four years {period}. {We emphasize again that Figs.\ref{fig:kp}-b,d show the whole Dilma-Temer term, though the division in groups is made before the impeachment and the alignment is calculated with the Dilma Rousseff's party (PT).}

Table~\ref{tab:k3} summarizes the size of each group $G_n^k$ and its alignment to the president's party $S_n^k$ for the eight presidential terms. We note that both the group size {\it and} its alignment needs to be relatively high to guarantee the president stability.
{One can see that Dilma I and Dilma II periods stand out by the small size of $G_1^3$ and large size of $G_3^3$ (though $G_2^3$ is also large in Dilma II). In this table we don't yet observe a small  value of $S_3^3$ for Dilma II period because the support of the $G_3^3$ only decays significantly very close to the impeachment process, as shown  in Fig.~\ref{fig:kp}. Even though Bolsonaro's $S_3^3$ is the only one below $0.6$, the size of $G_3^3$ is of only 9\% of the congress which makes it not really significant.}

\begin{table}
\caption{Absolute sizes of clusters $G_{i}^{3}$ (their proportion) and their support ${\cal S}_i^2$ to the president party in the correspondent period. }
\label{tab:k3}
\begin{tabular}{l|c|c|c||c|c|c}
\toprule
President & $G^{3}_{1}$ & $G^{3}_{2}$ & $G^{3}_{3}$ & $S^{3}_{1}$ & $S^{3}_{2}$ & $S^{3}_{3}$ \\  \hline \hline
\midrule
Collor    &      250(49\%) &      118(23\%) &      139(27\%) &              0.68 &              0.36 &              0.89 \\ 
Itamar    &      312(62\%) &       100(20\%) &       94(19\%) &               0.8 &              0.61 &              0.66 \\ \hline
FHC I      &      354(62\%) &       117(20\%) &      102(18\%) &              0.84 &               0.3 &              0.67 \\ \hline
FHC II      &      387(69\%) &      134(24\%) &       42(7\%) &               0.9 &              0.47 &              0.67 \\ \hline
Lula I     &      320(59\%) &      121(22\%) &      104(19\%) &              0.88 &              0.39 &              0.72 \\ \hline
Lula II     &       376(70\%) &      122(23\%) &       38(7\%) &               0.9 &              0.32 &              0.69 \\ \hline
Dilma I    &      138(25\%) &      115(21\%) &      300(54\%) &              0.89 &              0.45 &              0.76 \\ \hline
Dilma II    &       82(16\%) &      191(37\%) &      248(48\%) &              0.88 &               0.4 &              0.67 \\ 
Temer     &      363(68\%) &       107(20\%) &       67(12\%) &              0.88 &              0.27 &              0.66 \\ \hline
Bolsonaro &      370(72\%) &       101(20\%) &       45(9\%) &              0.87 &              0.23 &              0.54 \\ \hline
\bottomrule
\end{tabular}
\end{table}

We {analysed} carefully {the composition of the $G$ groups in order} to understand which parties form the group called  $G_3^3$ and we identify the following: i)  in Lula II, this group corresponds only to 7\% of the total number of congressmen and ii) it is composed by  congressmen of both groups $G_1^2$ and $G_2^2$. In other words, this small group $G_3^3$ is composed by congressmen from the effective opposition and effective {ruling coalition}. On the other hand, in the case of Dilma II, what happens is that there is a migration from the effective basis $G_1^2$ to this group $G_3^3$; the only parties that keep with the president are her own party and another one. Moreover, this new group is now composed by $48\%$ of the congressmen. If one adds this 45\% of congressmen to the previous 37\%  of congressmen identified in $G_2^2$ (which is the same as $G_2^3$), Dilma Rousseff had an effective opposition of almost the whole congress at the moment of her impeachment. The details about the migration of deputies between groups discriminated by parties are shown in Tables~\ref{tab:k2k3_migracoes_lula2} and \ref{tab:k2k3_migracoes_dilma2} of the Appendix \ref{append_migration} for two legislative terms.


\section{Summary and discussion}

Using data from all nominal roll calls occurred in the {Brazilian} Chamber of Deputies from 1991 to 2019, {after locating the roll calls in a cohesion phase space}, we computed how similar is the sequence of votes between two congressmen $\depi$ and $\depj$ and refer to  it as ``agreement between congressmen'', $\Aij$. Using this quantity as input data of the clustering algorithm  $k$-means, we separate  the congressmen in $k$ groups identified as $G_n^k$. For each of these groups, we defined  $S_n^k$ that quantifies how strong is each group's support to the president's party. We then measure how this quantity evolves in time for each legislative term.

From the assessment of the cohesion of the vote results in the roll calls, we observe that effective and party cohesion's seem to be correlated. This means that issues that unite the congressman from a same party are those that also unite the parties around the same cause and issues that divide the congressman are not pitching only parties with opposing ideas against each other, but also causing some kind of dissension within the parties themselves. This observation calls into question whether parties are an effective means for the electorate to associate its interests in a cohesive group of representatives. One does observe from the data, though, some kind of coherent behaviour in the congress in respect to the mater of forming a ruling coalition.

\begin{figure*}
\centering
\includegraphics[width=1.8\columnwidth]{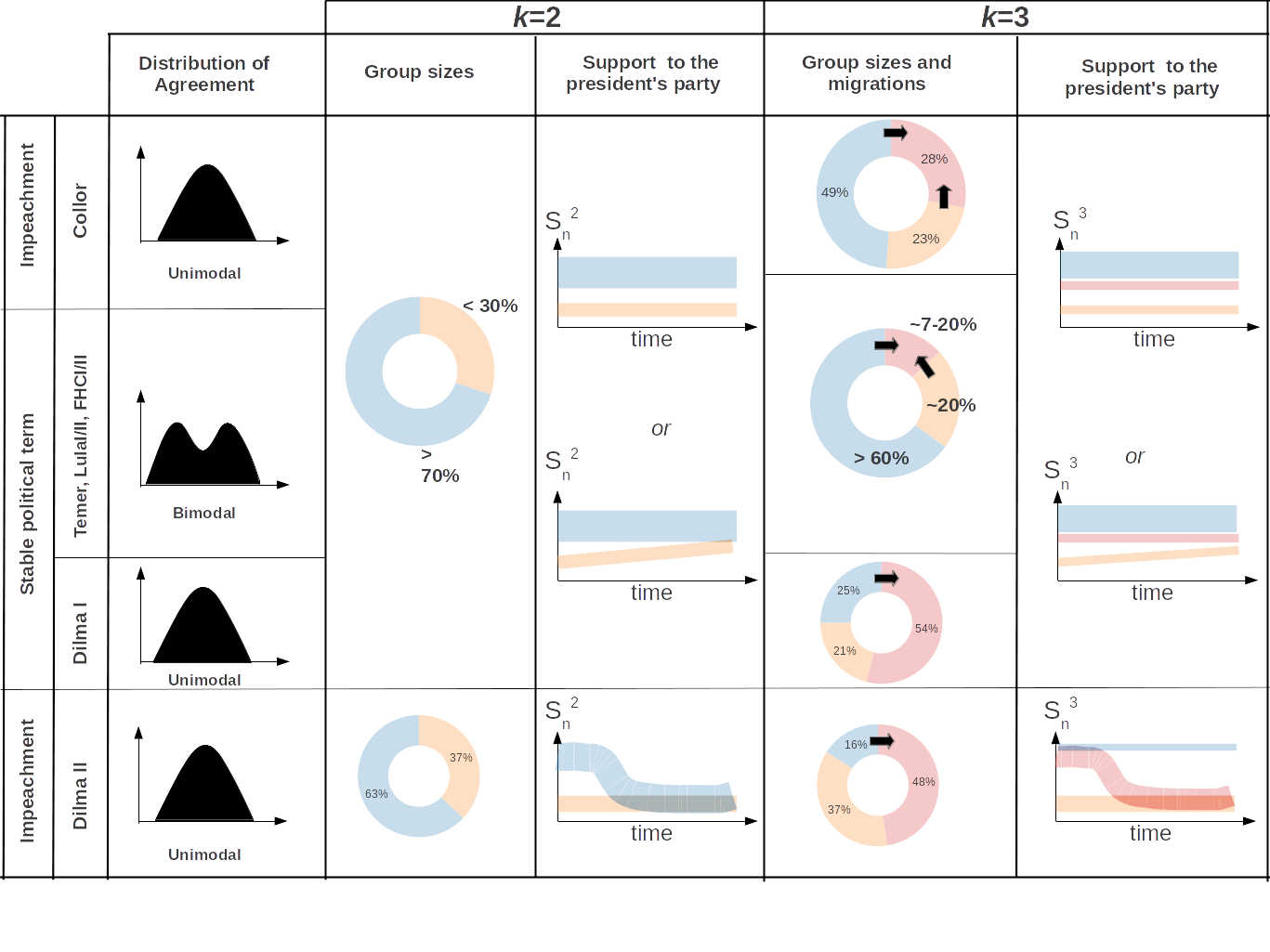}
\caption{Scheme of some characteristics of different legislative terms.  The first column shows the distribution of agreements  between votes of congressmen, indicating that for all stable periods the distribution is bimodal while for legislative terms with an impeachment and Dilma I (pre-impeachment) the behavior is unimodal. We then show the group sizes (represented by different colors in the donuts) and how the alignment to the president's party $S_n^k$ behaves along the legislative term for  the case where congressman are divided in $k=2$  and $k=3$ groups using the $k$-means algorithm.  The color code here is the same as for previous figures: blue for effective ruling coalition and orange for effective opposition and red for the third group. For $k=2$ we observe that the $G_2^2$ group is only bigger than $30\%$ in Dilma II. The support to president's party $S_n^k$ shows tree possible behavior: for all legislative terms except Dilma II, groups keep their support along time or the opposition group aligns more to the president's party close to the election. For DilmaII case, we observe that the ruling coalition support decays and reach the same support as the opposition. For $k=3$, besides showing the groups size, black arrows also indicate how congressman migrate to compose the  group $G_3^3$: for most legislative periods, $G_3^3$ is composed by congressman from two groups but for DilmaI and DilmaII, they completely migrate from the $G_1^2$, resulting in a very small effective ruling coalition. In the last column, it is shown how $S_n^k$ as a function of time. }
\label{fig:resumo}
\end{figure*}

Since 1991, Brazil has had eight direct presidential elections and in two of them there was an impeachment of the elected presidents. The periods without an impeachment are referred to as stable political terms in contrast with the terms with an impeachment, referred to as unstable. Our analyses show some differences between these two types of periods, as we now summarize and discuss with the help of the schema presented in Fig. \ref{fig:resumo}.

Fig. \ref{fig:resumo} is divided by lines and rows. In lines we show the legilastive terms  almost in cronological order.  The exception is Temer, who governed after Dilma but we aggregate him with other presidents with political stability. Bolsonaro is not represented in this figure  because his government is only at its second year. We will comment about him later based on what we can observe now. In rows we show different types of measures presented in this work.

Stable political periods can be characterized as follows. The distribution of $\Aij$ presents two peaks, while a continuous, unimodal distribution is observed in legislative periods that have had an impeachment. The exception is Dilma I, where we observe a unimodality in this distribution. Although she was not impeached in this period, it  might mean that the instability that lead to her impeachment in the second term had already began in her first term. When congressmen are  divided in $k=2$ groups called $G_1^2$ and $G_2^2$, it is observed that i) $G_1^2$ is composed by at least $70\%$ of the congressmen, ii) it presents a high support for the president's party and iii) this support is high along the whole legislative term. The biggest group with a high support to the president's party is associated with an effective ruling coalition, which correlates well with a self-declared coalition in cases where this comparison is possible. All this suggests that the stability of the presidents requires a big and  cohesive effective coalition during the whole mandate. Our data show that ``big'' means at least 70\% of the congressmen, although we do not have a theory to assume that this is a minimum value. Moreover, we analysed the stability of this division in two groups by dividing them in $k=3$. We observe that, for stable legislative terms,  the third group is very small (from 7\% to 20\% of the congressmen), it is composed by congressmen from the two other groups  $G_1^2$ and $G_2^2$  and presents a high value of support to the president's party during the whole legislative term. In these cases, the third group is effectively part of a coalition with the government.

Dilma's impeachment in her second term is characterized by the decay of the $S_1^2$ along the legislative term. When the congress is separated into 3 groups, one clearly observes that this coalition represented by the group  $G_1^2$ which is composed by 63\% of congressmen splits into two groups: a very small group $G_1^3$ with around 16\% of the congressmen that keeps its support and the other part that aligns with the opposition and the group $G_3^3$ with $37\%$ of the congressman. These 16\% of congressmen corresponds to her own party and another one; all other parties, including PMDB which was the party of the vice-president migrate to an opposition. Collor was also impeached and does not present a decay of $S_1^2$ but the division by tree groups also shows that his real coalition was smaller than 50\% of the congress, which suggests that the division into two effective groups with a supportive and cohesive coalition was not stable.


\section{Conclusion}

Thirty years ago the concept of  coalisional presidentialism was proposed~\cite{abranches1988} to understand presidential regimes with many parties. The idea is simple: if a president cannot form a majority in the Parliament with his/her own party, a coalition between many parties is necessary. Moreover, it was also shown that the projects proposed by the president's party have a very high rate of approval in the Brazilian congress \cite{LIMONGI2006}. These results are in line with our observations during stable periods: Indeed we observe that stable periods have an important proportion of the congressmen in the ruling coalition with the president and the support of this group is very high. Our data also show that, once this coalition breaks down, the president's mandate does not survive. We are not able to say anything about why this happens but it can perhaps help to predict if this instability will happen with a given president. Dilma Rousseff's first mandate was signaling this instability as can be seen in some of our quantities as summarized in Tab.~\ref{tab:k3}. And if these quantities measured here have any prediction power,  based on the first year of Bolsonaro's government, we would say that {it was} a stable political period. We remind that these analyses are until December 2019 and that we are right now living an important pandemic crisis that may destabilize the system. 

It would be important for next studies to build statistical models based on some of the characteristics of the congress that could predict some properties of each term that could lead to an instability and also to compare with results from roll calls in other countries. For example, the minimum coalition size and the minimal support to maintain the mandate. In this respect, we observe that the measure of cohesion proposed in this work, Fig. \ref{fig:votesinps}, shows that it is possible
to classify the roll calls in two separate groups, one very cohesive that align all parties together and, therefore, does not seem to bring relevant information to discriminate groups and another where the dissension and dynamics pitching the congressmen with different opposing ideas or strategies is happening. This observation also call in question the roll of the exaggerated number of parties in the Brazilian political scene. The above mentioned measure of cohesion  can be used as a filter of the roll calls that are relevant to separate parties with real different ideologies. It could then help building an objective measure of how many parties would be necessary to represent all the ideologies present in the Brazilian national Congress.


\appendix


\section{Criterion of exclusion of congressmen with low legislative activity}
\label{append_a}

There are congressmen that, for a variety of reasons, such as becoming ministers or taking another office appointed by the president or governors, have very few roll calls where they voted differently than Absent. Those congressmen usually have some very high (or very low) agreements with other congressmen with very different (or very similar) voting records, which introduces noise in the analysis. 

\begin{figure}[h!]
\centering
\includegraphics[width=0.95\columnwidth]{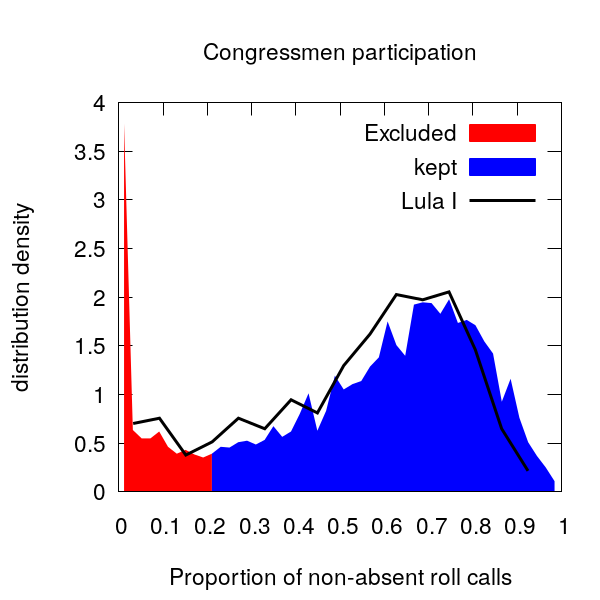}
\caption{Distribution of the number of non absent votes for the congressmen in all periods. In red are the excluded congressmen, in blue the kept ones. Lula I mandate is shown individually for comparison. }
\label{fig:explicaExclusaoDeputados}
\end{figure}

{To reduce this source of noise, we analyze the number of times that each congressman voted in the roll calls in a given legislative term. The distribution of it is shown in  Fig.\ref{fig:explicaExclusaoDeputados} for all legislative terms together and for one particular term, showing that there is a regular pattern in this measure. We observe a peak close to zero that correspond to congressman who rarely vote and  a second peak at higher values in this distribution. We then introduce a cutoff to exclude congressman who voted in less than $20\%$ of the roll calls. We have played with this cutoff value and  our results are robust against these small changes. } 
In red are the excluded congressmen, in blue the kept ones. 
{By choosing a 20\% cutoff the largest number of excluded congressmen happens in the Itamar term when 15.2\% of them were excluded. }

As there are substitutes for congressmen that for the many reasons leave the mandate and we consider each congressman individually, the total number of congressmen in our analyses for a given term can be greater than 513, which is the actual number of seats in the congress.

\section{Details about the migration of the congressmen between groups discriminated by parties}
\label{append_migration}

In Sec.~\ref{sec:timeevolution} we analyzed the separation of congressmen in two or three clusters and their support $S_i^k$ to the president's party. In this appendix we detail that analysis.

In Tables~\ref{tab:k2k3_migracoes_lula2}, \ref{tab:k2k3_migracoes_dilma2} we show the crosstabulation of parties and clusters found by \emph{k}-means for $k=2$ and $k=3$ with their absolute and relative numbers for Lula II and Dilma II periods, respectively.

We remark that parties belong mostly to one cluster only. That is, the boundary between clusters respect the boundaries between parties. 

Also, in Lula II, $G_3^3$ is small, with only 7\% of the congressmen, and the only party that is largely in it is PSOL. In Dilma II, the only parties that remain in $G_1^3$ are PT (Dilma's own party) and PCdoB, while $G_2^3$ is essentially equal to $G_2^2$, which means that the government coalition is fractured.

\begin{table*}
\caption{Crosstabulation of parties and clusters found by \emph{k}-means for $k=2$ and $k=3$ with their absolute and relative numbers for Lula II period. When there are more than one party name in the same line means that the party changed its name.}
\label{tab:k2k3_migracoes_lula2}
\centering
\begin{tabular}{l||c|c||c|c|c}
Party & $G_1^2$ & $G_2^2$  & $G_1^3$ & $G_2^3$ & $G_3^3$ \\  \hline \hline 
DEM/PFL & 11 (17\%) & 55 (83\%) & 8 (12\%) & 54 (82\%) & 4 (6\%)\\ 
PAN & 5 (100\%) & 0 (0\%) & 5 (100\%) & 0 (0\%) & 0 (0\%)\\ 
PCdoB & 13 (100\%) & 0 (0\%) & 13 (100\%) & 0 (0\%) & 0 (0\%)\\ 
PDT & 25 (100\%) & 0 (0\%) & 23 (92\%) & 0 (0\%) & 2 (8\%)\\ 
PHS & 2 (100\%) & 0 (0\%) & 2 (100\%) & 0 (0\%) & 0 (0\%)\\ 
PMDB/MDB & 89 (97\%) & 3 (3\%) & 82 (89\%) & 1 (1\%) & 9 (10\%)\\ 
PMN & 5 (100\%) & 0 (0\%) & 5 (100\%) & 0 (0\%) & 0 (0\%)\\ 
PP/PPB & 40 (98\%) & 1 (2\%) & 40 (98\%) & 0 (0\%) & 1 (2\%)\\ 
CIDADANIA/PPS & 5 (28\%) & 12 (71\%) & 4 (24\%) & 10 (59\%) & 3 (18\%)\\ 
PL/PR & 36 (100\%) & 0 (0\%) & 36 (100\%) & 0 (0\%) & 0 (0\%)\\ 
REPUBLICANOS/PRB/PMR & 1 (100\%) & 0 (0\%) & 1 (100\%) & 0 (0\%) & 0 (0\%)\\ 
PSB & 30 (100\%) & 0 (0\%) & 25 (83\%) & 0 (0\%) & 5 (17\%)\\ 
PSC & 6 (100\%) & 0 (0\%) & 6 (100\%) & 0 (0\%) & 0 (0\%)\\ 
PSDB & 6 (10\%) & 55 (90\%) & 5 (8\%) & 55 (90\%) & 1 (2\%)\\ 
PSOL & 0 (0\%) & 3 (100\%) & 0 (0\%) & 0 (0\%) & 3 (100\%)\\ 
PT & 86 (99\%) & 1 (1\%) & 82 (94\%) & 1 (1\%) & 4 (5\%)\\ 
PTB & 26 (96\%) & 1 (4\%) & 22 (81\%) & 0 (0\%) & 5 (19\%)\\ 
PTC/PRN & 3 (75\%) & 1 (25\%) & 3 (75\%) & 0 (0\%) & 1 (25\%)\\ 
AVANTE/PTdoB & 1 (100\%) & 0 (0\%) & 1 (100\%) & 0 (0\%) & 0 (0\%)\\ 
PV & 13 (93\%) & 1 (7\%) & 13 (93\%) & 1 (7\%) & 0 (0\%)\\ \hline \hline  
Total & 403 (75\%) & 133 (25\%) & 376 (70\%) & 122 (23\%) & 38 (7\%)\\ \hline
\end{tabular}
\end{table*}

\begin{table*}
    \caption{Crosstabulation of parties and clusters found by \emph{k}-means for $k=2$ and $k=3$ with their absolute and relative numbers for Dilma II period. We highlight in the table the only two parties that stayed in the group $G_1^3$, which is the ruling coalition.}
    \label{tab:k2k3_migracoes_dilma2}
    \centering
\begin{tabular}{l||c|c||c|c|c}
Party & $G_1^2$ & $G_2^2$  & $G_1^3$ & $G_2^3$ & $G_3^3$  \\ \hline \hline 
DEM/PFL & 0 (0\%) & 22 (100\%) & 0 (0\%) & 22 (100\%) & 0 (0\%)\\ 
{\bf PCdoB} & {\bf 13 (100\%)} & {\bf 0 (0\%)} & {\bf 13 (100\%)} & {\bf 0 (0\%)} & {\bf 0 (0\%)}\\ 
PDT & 18 (90\%) & 2 (10\%) & 1 (5\%) & 2 (10\%) & 17 (85\%)\\ 
PATRIOTA/PEN & 2 (100\%) & 0 (0\%) & 0 (0\%) & 0 (0\%) & 2 (100\%)\\ 
PHS & 5 (100\%) & 0 (0\%) & 0 (0\%) & 0 (0\%) & 5 (100\%)\\ 
PMDB/MDB & 60 (88\%) & 8 (12\%) & 0 (0\%) & 6 (9\%) & 62 (91\%)\\ 
PMN & 2 (67\%) & 1 (33\%) & 0 (0\%) & 1 (33\%) & 2 (67\%)\\ 
PP/PPB & 26 (65\%) & 14 (35\%) & 0 (0\%) & 14 (35\%) & 26 (65\%)\\ 
CIDADANIA/PPS & 0 (0\%) & 11 (100\%) & 0 (0\%) & 11 (100\%) & 0 (0\%)\\ 
PL/PR & 33 (97\%) & 1 (3\%) & 1 (3\%) & 1 (3\%) & 32 (94\%)\\ 
REPUBLICANOS/PRB/PMR & 20 (100\%) & 0 (0\%) & 0 (0\%) & 0 (0\%) & 20 (100\%)\\ 
PROS & 12 (86\%) & 2 (14\%) & 0 (0\%) & 2 (14\%) & 12 (86\%)\\ 
PRP & 3 (100\%) & 0 (0\%) & 0 (0\%) & 0 (0\%) & 3 (100\%)\\ 
PRTB & 1 (100\%) & 0 (0\%) & 0 (0\%) & 0 (0\%) & 1 (100\%)\\ 
PSB & 2 (6\%) & 31 (94\%) & 0 (0\%) & 31 (94\%) & 2 (6\%)\\ 
PSC & 2 (15\%) & 11 (85\%) & 1 (8\%) & 9 (69\%) & 3 (23\%)\\ 
PSD & 29 (85\%) & 5 (15\%) & 0 (0\%) & 5 (15\%) & 29 (85\%)\\ 
PSDB & 0 (0\%) & 53 (100\%) & 0 (0\%) & 53 (100\%) & 0 (0\%)\\ 
DC/PSDC & 2 (100\%) & 0 (0\%) & 0 (0\%) & 0 (0\%) & 2 (100\%)\\ 
PSL & 1 (100\%) & 0 (0\%) & 0 (0\%) & 0 (0\%) & 1 (100\%)\\ 
PSOL & 0 (0\%) & 5 (100\%) & 0 (0\%) & 5 (100\%) & 0 (0\%)\\ 
{\bf PT} & {\bf64} {\bf(98\%)} & {\bf1 (2\%)} & {\bf64 (98\%)} & {\bf1 (2\%)} & {\bf0 (0\%)}\\ 
PTB & 19 (76\%) & 6 (24\%) & 0 (0\%) & 6 (24\%) & 19 (76\%)\\ 
PTC/PRN & 2 (100\%) & 0 (0\%) & 0 (0\%) & 0 (0\%) & 2 (100\%)\\ 
PODEMOS/PTN & 4 (100\%) & 0 (0\%) & 1 (25\%) & 0 (0\%) & 3 (75\%)\\ 
AVANTE/PTdoB & 2 (100\%) & 0 (0\%) & 0 (0\%) & 0 (0\%) & 2 (100\%)\\ 
PV & 3 (33\%) & 6 (67\%) & 1 (11\%) & 6 (67\%) & 2 (22\%)\\ 
SD & 1 (6\%) & 16 (94\%) & 0 (0\%) & 16 (94\%) & 1 (6\%)\\ \hline \hline  
Total & 326 (63\%) & 195 (37\%) & 82 (16\%) & 191 (37\%) & 248 (48\%)\\ \hline
    \end{tabular}
\end{table*}


\bibliographystyle{apsrev4-1}
\bibliography{congressoNacional}

\end{document}